\documentclass[aps,reprint,twocolumn,amsmath,amssymb,nofootinbib]{revtex4-1}

\usepackage{xr}
\usepackage{fullpage}
\usepackage{graphicx,color}
\usepackage[dvipsnames]{xcolor}
\usepackage{makecell}
\usepackage{multirow}
\usepackage[normalem]{ulem}
\usepackage{subcaption}
\usepackage{hhline}
\usepackage[symbol]{footmisc}

\usepackage{booktabs}
\def\RR{{\mathbf R}}

\def\N{{\mathbb N}}

\def\R{{\mathbb R}}

\def\E{E}

\def\1{{\mathds{1}}}
\def\RI{{\rm RI}}
\def\RPI{{\rm RPI}}

\def\bR{{\mathbf r}}

\def\bX{\mathbf{x}}
\def\If{f}
\def\Ip{p}
\def\Is{s}

\def\x{u}

\def\Ee{{\cal E}}
\def\Fe{{\cal F}}
\def\Ve{{\cal V}}

\def\Er{E}
\def\Erp{\Phi}

\newcommand\blfootnote[1]{%
  \begingroup
  \renewcommand\thefootnote{}\footnote{#1}%
  \addtocounter{footnote}{-1}%
  \endgroup
}

\begin{document}

\title{Regularised Atomic Body-Ordered Permutation-Invariant Polynomials \\
for the Construction of Interatomic Potentials}

\author{Cas van der Oord$^*$ \thanks{\texttt{casv2@eng.cam.ac.uk}}}
\affiliation{Engineering Laboratory,
University of Cambridge,
Trumpington Street,
Cambridge, CB2 1PZ,
United Kingdom}

\author{Genevi\`eve Dusson$^*$\thanks{\texttt{g.dusson@warwick.ac.uk}}}
\affiliation{Mathematics Institute, University of Warwick, Coventry CV47AL, United Kingdom}

\author{G\'abor Cs\'anyi\thanks{\texttt{gc121@cam.ac.uk}}}
\affiliation{Engineering Laboratory,
University of Cambridge,
Trumpington Street,
Cambridge, CB2 1PZ,
United Kingdom}

\author{Christoph Ortner\thanks{\texttt{c.ortner@warwick.ac.uk}}}
\affiliation{Mathematics Institute, University of Warwick, Coventry CV47AL, United Kingdom}

\date{\today}

\blfootnote{$^*$these authors contributed equally to the work}

\begin{abstract}
We investigate the use of invariant polynomials in the construction of data-driven
interatomic potentials for material systems. The ``atomic body-ordered
permutation-invariant polynomials'' (aPIPs) comprise a systematic basis and are  constructed to preserve the symmetry of
the potential energy function with respect to rotations and
permutations. In contrast to kernel based and artificial neural
network models, the explicit decomposition of the total energy as a sum of
atomic body-ordered terms allows to keep the dimensionality of the
fit reasonably low, up to just 10 for the 5-body
terms. The {\em explainability} of the potential is aided by this
decomposition, as the low body-order components can be studied and
interpreted independently.
Moreover, although polynomial basis functions are thought to extrapolate poorly,
we show that the low dimensionality combined with careful regularisation actually leads
to better transferability than the high dimensional, kernel based Gaussian
Approximation Potential.
\end{abstract}


\maketitle

\section{Introduction}

There is a long and successful history of using empirical interatomic
potentials for the simulation of materials~\cite{Finnis:2004}. One
approach is to treat such models as purely phenomenological, setting
out a few key features of the true interaction to be captured
(e.g. the stability ordering of certain phases), and investigate what
other properties, both macroscopic or indeed microscopic, follow from
these. More recently, as electronic structure calculations,
particularly density functional theory (DFT)~\cite{DFTbook:1980}, has
increased both in accuracy and availability, there has been a widely
shared desire for potentials to match the Born-Oppenheimer potential
energy surface as closely as
possible~\cite{ercolessi1994interatomic,murrell1984molecular}. This change of attitude
in the materials simulation community has come rather later than the analogous one in
the world of organic force fields, partly due to the more systematic
nature of quantum chemistry methods applicable to small
molecules~\cite{atkins2011molecular, huang:2005}.

Already a decade ago, it was clear  that
empirical potentials had reached their limits in terms of their
ability to match the potential energy surface of DFT, essentially due
to the use of simple, physically interpretable functional forms. At
around the same time significant developments started in which models
with thousands of free parameters (so-called {\em high-dimensional}
models) were fitted to electronic structure data. The methods are
borrowed from machine learning, e.g. artificial neural networks
(ANN)~\cite{ANN:Jain:1996, Bishop:2006:PRM:1162264, Behler2007-ng} and
Gaussian processes (GP)~\cite{Rasmussen:2005:GPM:1162254,
  Bartok2015-xc}.  Although formally these models contain many degrees
of freedom, they are often called {\em non-parametric}, because there
are either good recipes for determining the best parameters that fit
the data (in the case of training ANNs), or linear algebra expressions
in the case of GPs. The few model parameters that are still adjusted
by hand or by other ad-hoc recipes are called {\em hyper-parameters},
e.g. the nonlinear transfer function of ANN units, or the kernel
shapes in GPs. It was understood early on, similarly to the more
traditional applications of machine learning, that it is advantageous
to use an appropriate representation of the input data (atomic
positions in this case) that captures all the known symmetries present in the
problem~\cite{Bartok:2013cs}. These models achieve very high
accuracy on the training datasets and, when carefully used, on
configurations that are ``near'' the training, e.g., in a molecular
dynamics run under similar conditions. However, the transferability of
such models can still be poor. A recent attempt at assembling and fitting to a
very large and diverse training dataset of elemental
silicon~\cite{Bartok:2018ih}, while generally staying physically sensible away
from the training data, still showed up to 20\% error in formation energies and
migration barriers of some defects that were not in the training set. The data
requirements to achieve even this level of transferability are expected to grow
significantly for multicomponent systems.

While a better choice of representations and kernel functions may
improve the transferability somewhat, it is conceivable that
high-dimensional fits will, by their very nature, always suffer from
this problem.

Similar effects can be seen in a related field, the fitting of the
potential energy surfaces of  molecules to high level, wave function based
quantum chemistry calculations. This endeavour has a rich
history~\cite{Partridge:1997kh, Bramley:1993ik, Szalewicz:2008, Xie:2005,
Babin:2013fs, Qu:2018gj}, which also includes high-dimensional nonparametric
fits that are very accurate for small systems (a handful of atoms), yet
it is recognised that once the dimensionality reaches a few tens, the fitting
task becomes extremely difficult.

At present, the only plausible way to break the curse of
dimensionality is to explicitly or implicitly identify low-dimensional
structures of the potential energy surface. If this can be done
explicitly then the energy  can be broken up into multiple
low-dimensional terms, ideally ensuring that higher dimensional terms
account for less variation. The challenge is to do this
generally, systematically, and without sacrificing accuracy.

A time-tested and obvious way to introduce low dimensional terms is to
use the {\em body order expansion} applied in an atom-by-atom
fashion~\cite{Griebel2005-fx,Rabitz1999-wj,Stillinger1985-ul,Nguyen2018-le,Glielmo2018-ti,Sharma2006-le},
i.e. define the total energy as a sum of one-atom, two-atom (pair),
three-atom (angle) terms, and so forth.
Let $\RR \equiv \{\bR_j\}_{j=1}^M$ be the
positions of $M$ atomic nuclei of the same species, representing a {\em configuration}\ of atoms (perhaps but not necessarily with periodic boundary conditions), then we write the total energy as
%
\begin{align}
   \notag
   \E(\RR) &= \sum_{j} \E_1(\bR_j) \\
   \notag
      & \qquad + \frac12 \sum_{i \neq j} \E_2(\bR_i, \bR_j) \\
\label{eq:bo-expansion}
      & \qquad + \frac{1}{3!} \sum_{i \neq j \neq k} \E_3(\bR_i, \bR_j, \bR_k) \\
      \notag
      & \qquad + \cdots +  \\
      \notag
      & \qquad + \frac{1}{N!} \sum_{i_1 \neq \dots \neq i_N} \E_N\big(\bR_{i_1}, \dots, \bR_{i_N}).
\end{align}
Note that if the $n$-body function $\E_n$ is permutation invariant then $\frac{1}{n!} \sum_{i_1 \neq \dots \neq i_n}$
may of course be rewritten as $\sum_{i_1 < \dots < i_n}$. Periodic boundary conditions are treated by taking into account the periodic images of atoms within the computational cell in \eqref{eq:bo-expansion}.

We must strongly emphasize the distinction, on the one hand, between such a
decomposition of the total energy into additive components, each of which only
depends on few coordinates, and on the other hand the construction of {\em
complete, many-body, invariant representations} of $\RR$ using various combinations of two-
and three-body functions, which are subsequently used in a single
high-dimensional nonlinear nonparametric
fit~\cite{rupp2012,huo2017unified,Behler:2017}. For example, consider the
symmetry functions of Behler and Parrinello~\cite{Behler:2011it}. Although each
element of the descriptor is itself built out of just interatomic distances
(2-body) or angles (3-body), the entire descriptor vector taken as a whole is a
high dimensional description of the neighbour environment of an atom, and
subsequent fits are of functions in that high dimensional space. In contrast,
the individual energy terms in \eqref{eq:bo-expansion} are all low dimensional,
and it is to be expected that they can be fitted using much less data, since low
dimensional spaces can be covered comprehensively by a relatively small number
of training configurations. In particular we conjecture excellent extrapolation
properties.  Note that this does not require an{\em\ a priori}\ definition or
calculation of these terms, the fitting is still to be made to total energy $\E$
and its derivatives (forces and stresses) corresponding to configurations with many atoms.

The utility of additive body order expansions has been recognised in
the context of the recent machine learning based potentials. By
writing the total energy as a sum of pair, triplet and a many-body
terms with explicit weight
factors~\cite{Bartok2015-xc,Deringer:2017ea,Bartok:2018ih}, it is
possible to prevent catastrophically erroneous predictions at small
interatomic distances. Note that the failure of high dimensional fits
at small interatomic distances is well known by the quantum chemists
who fit small molecule potential energy surfaces~\cite{Nandi2019-bl}.

However, merely writing the total energy as a body order expansion
does not in itself bring the benefits of low dimensionality. The terms
in \eqref{eq:bo-expansion} may be highly redundant, e.g. a general
three-body potential includes all possible two-body potentials by
simply not depending on one of its arguments, and this is true for all
orders. A potential fitting methodology whose terms are intrinsically
body-ordered was introduced as the Moment Tensor
Potentials (MTPs)~\cite{Shapeev:2016kn}, although it would appear that no
explicit use is made of the lowest dimensional terms to maximise
transferability. Recently, the Atomic Cluster Expansion (ACE) was
introduced and the connection between the body order expansion and the
high dimensional representations used in the earlier many-body fits
was also made formally explicit~\cite{Drautz:2019}. There, transferability
is achieved by defining and calculating the low order terms in
\eqref{eq:bo-expansion} explicitly using the total energy function of
small clusters: the order $M$ term is the interaction energy of the
$M$-atom cluster in vacuum (with lower order interactions subtracted).

In both the MTP and the ACE approaches, a basis of symmetric polynomials is introduced, whose elements
are body-ordered, and a linear fit in this basis is the fundamental modelling
tool. The high computational cost of the many-body expansion is avoided by
taking the entire many-body environment of each atom as a spatial density, and
projecting it onto a rotationally invariant basis set, the components of which
turn out to be body-ordered. This density trick is the same that is
used by high-dimensional descriptors, including
SOAP-GAP~\cite{Bartok:2013cs}, Behler-Parrinello symmetry
functions~\cite{Behler:2011it}, and the bispectrum~\cite{bartok2010gaussian,THOMPSON2015316}, but without taking advantage of the
body-ordered decomposition.

Finally, it is notable that yet another route to the atomic body ordered
expansion is afforded by the Generalised Pseudopotential Theory
approach of Moriarty~\cite{Moriarty:1977, Moriarty:1982, Moriarty:1988}. GPT treats
perturbations of the electron density within the framework of density
functional theory using the uniform electron gas as reference, and via
a series of approximations derives individual body ordered terms
formally - but also including the unit cell volume as a variable. It has had considerable success in modelling single species
defect-free metals in an essentially parameter-free manner. Its
extension to more complex materials is not straightforward, and
calculating high body order terms (3- and 4-body terms) is rather
complicated even for the simpler cases.  Nevertheless GPT can be
thought of as the explanation (or {\em justification}) of why the body
order expansion is a good idea for strongly bound condensed phase
materials, especially metals.

In this work, we will consider each body ordered term as an independent function
to be fitted, but only whose sum is known. We will define a basis set with which
we do a linear fit, so at the end, all the unknown basis coefficients may be
determined in a single linear least squares problem, but separate and distinct
distance based cutoffs and regularisation strategies are applied to each term.
We will also forgo the vacuum-cluster definition of each term, and instead fit
the basis coefficients directly to condensed phase data only. The advantage in
fitting the body-order expansion to a condensed phase training set is that it
can be expected to converge faster. Indeed, there is significant empirical
evidence for this, such as the relative success of few-body interatomic
potentials, cluster expansion for alloys~\cite{nelson2013cluster},
GPT~\cite{Moriarty:1977, Moriarty:1982, Moriarty:1988}, however, we are not
aware of any rigorous results that explicitly show this in general.

For defining a basis set for the body ordered terms, we employ the
theory of {\em permutationally invariant polynomials} (PIP) a
technique based on classical invariant theory, introduced to molecular
modelling for fitting the potential energy surfaces of small
molecules; see e.g.~\cite{Braams2009-wi,Babin:2013fs,Nguyen2018-le} and references
therein. To adapt this formalism for condensed phase covalent
materials we make two modifications: (i) an explicit introduction of
distance-based cutoffs into the basis functions, and (ii) each body-ordered term
has its own set of PIPs, taking account of the specific symmetry group of that
term. By contrast, in the original application of PIPs to small molecules, each
molecule (or a set of small molecules taken as a cluster) had its potential
energy surface defined and fit with the appropriate set of PIPs and each new molecule or molecular cluster required a completely new fit.
 In a somewhat similar vein to our ideas here, the possibility to apply PIPs to manually determined subsets (fragments) of a
molecule has previously been suggested in~\cite{Braams2009-wi} and very recently first explored in~\cite{Qu2019-vb}.

We will call our basis set ``atomic PIPs'' or aPIPs, to emphasize
that the body order expansion inherent to the use of PIPs is done here on
an atomic rather than molecular basis.
For the sake of simplicity, we limit the exposition here to elemental
materials, but the formalism generalises naturally to multiple
species, exactly in the same way as the original PIP formalism does for
molecules.

Our overarching goal, for which the aPIP fits serve as examples, is to demonstrate for the case of materials (rather than isolated molecules) that (i)  low body order potentials can reach the same high accuracies that high dimensional fits can, and (ii) polynomial fits can be used to improve generalisation properties of interatomic potentials and that sophisticated regularisation is the key to achieving this.

\section{Symmetric Polynomial Basis}
\label{sec:pots}
Our starting {\em assumption} is that \eqref{eq:bo-expansion} can be used to construct
high-accuracy PESs for moderate to low body-order $N$, which is
certainly the case empirically. Next, we require that the individual
contributions $\E_n$ inherit rotation and permutation invariance (RPI)
from $\E$. Our aim is then to construct systematically improvable
functional forms to represent individual  $\E_n$ functions that exactly preserve these
symmetries:
\begin{enumerate}
   \item Construct coordinate systems
   \[
       \bX_n = \bX_n(\bR_1, \dots, \bR_n)
    \] that are
   continuous, rotation and permutation invariant, and represent
   \[ E_n(\{\bR_i\}_{i=1}^n) =
   \Erp_n(\bX_n(\{\bR_i\})). \]
   \item Choose basis sets $\{ B_{nj} \}$ and represent
   \[ \Erp_n(\bX) = \sum_{j}
      c_{nj} B_{nj}(\bX). \]
   \item Apply a cut-off mechanism to prevent inclusion  of clusters with atoms that are very far from each other that have negligible contributions to the total energy.
   \item Use regularised linear least squares (i.e., ridge regression, force matching~\cite{ercolessi1994interatomic}) to
   determine the coefficients $\{c_{nj}\}_{n,j}$, using total energies, forces and stresses calculated by a first principles electronic structure approach as training data.
\end{enumerate}

Several aspects of this strategy are familiar from recent
machine-learning models: For example, the motivation for employing a
RPI coordinate system is the same as for the use of symmetry functions
descriptors for ANN potentials of~\citet{Behler2007-ng} as well as the
SOAP descriptor and kernel of the GAP framework of~\citet{Bartok2015-xc}.
Employing a polynomial basis to obtain a systematically improvable
functional form was also proposed by~\citet{Braams2009-wi, Shapeev2016-hn}. We will
import much of the invariant theory methodology for the representation
of symmetric polynomials from~\citet{Braams2009-wi}.  In the following
we will show how this approach leads to a large design space and in
particular describe different coordinate systems as well as the basis
functions to represent the potential energy.

\subsection{Distance-based potentials}
\label{sec:pots:BL}
Our first construction of a RPI coordinate system is achieved by closely
following the ideas of~\citet{Braams2009-wi}: we first choose rotation
invariant ($\RI$) coordinates of transformed interatomic distances,
\[
   \E_n(\{\bR_i\}_{i=1}^n)
   =
   \Er_n^{\rm D}(\{\x_{ij}\}_{i<j = 1}^n),
\]
where $u_{ij}$ denotes a {\em distance transform} such as
euclidean distance itself or the common Morse or inverse distance
variables,
\[
\x_{ij} = r_{ij}, \quad
u_{ij} = e^{-\alpha r_{ij}}, \quad u_{ij}=r_{ij}^{-p},
\]
where $\alpha, p > 0$. The superscript ``D'' in $\Er_n^{\rm D}$ indicates that the
arguments of the function are distances. Alternatively, distances and angles can be used, which we discuss below and will denote with the superscript ``DA''.

In the case of the 2-body term, $\x_{12}$ is already a RPI coordinate system.
For the 3-body contribution, the permutation invariance of $\E_3$ with respect
to exchanging atom indices gives rise to full $S_3$ permutation invariance of
$\Er_3^{\rm D}$. Here, and throughout, $S_n$ denotes the symmetric group over $n$
elements. The elementary symmetric polynomials therefore give a permutation
invariant coordinate system,
\begin{align*}
   \E_3(\bR_1, \bR_2, \bR_3)
   &= \Er_3^{\rm D}(\x_{12}, \x_{13}, \x_{23}) \\
   &= \Erp_3^{\rm D}(f_{3,1}, f_{3,2}, f_{3,3}),
\end{align*}
where
\begin{align}
   f_{3,1} & = \x_{12}+\x_{13}+\x_{23}, \nonumber \\
   f_{3,2} & = \x_{12} \x_{13}+\x_{12} \x_{23}+\x_{13} \x_{23},\nonumber \\
   f_{3,3} & = \x_{12} \x_{13} \x_{23}. \label{eq:I3}
\end{align}
The key property of the coordinates $f_3 = (f_{3,1}, f_{3,2}, f_{3,3})$ is that they completely
define $(\bR_1, \bR_2, \bR_3)$ up to rotations and permutations. The choice of
such invariant coordinates is not unique. For example, we may choose $\If_n
= \sum_{i<j} u_{ij}^n, n = 1,2,3$ and in \S~\ref{sec:bapots} we replace distance
coordinates with distance and angle coordinates.

\begin{figure}
   \includegraphics[scale = 0.5]{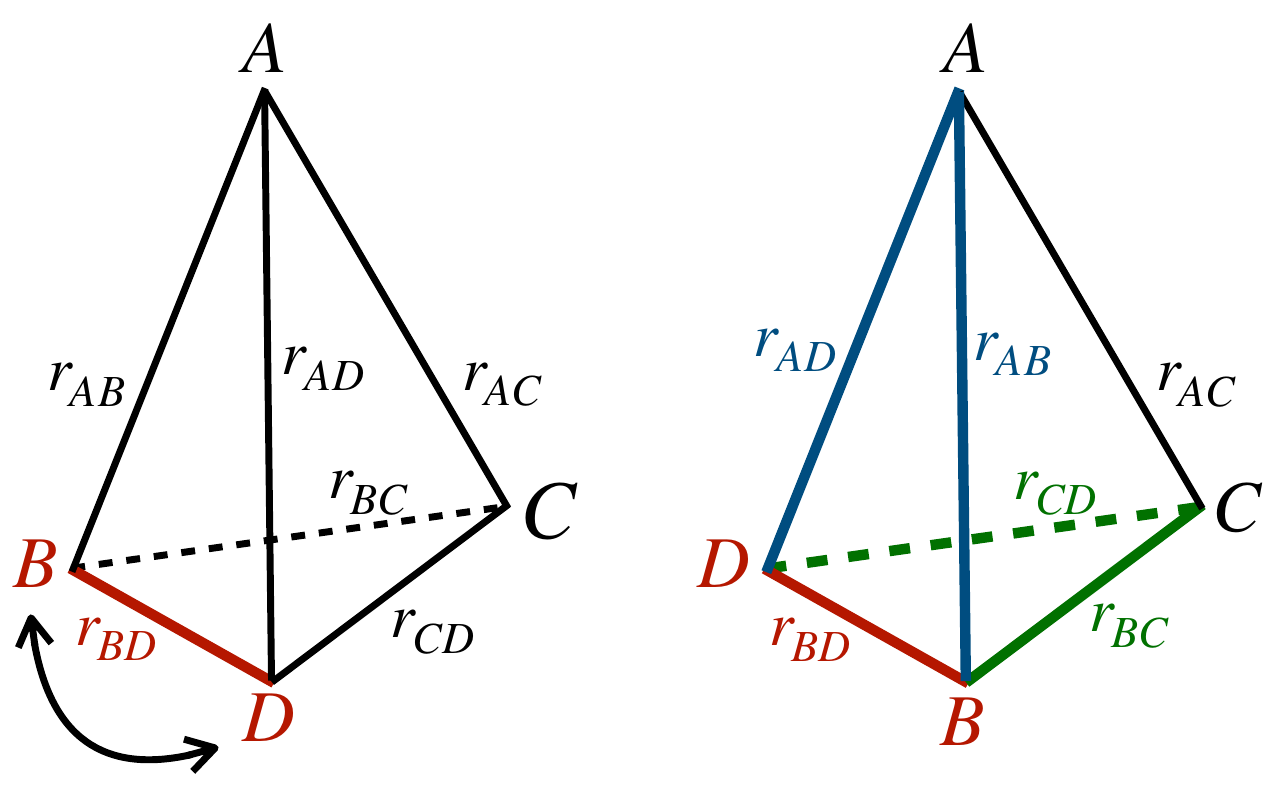}
   \caption{Permutation of the edges of a tetrahedron when two corners are permuted. In blue and green are permuted edges. Although there are six edges, the symmetry group of the edges induced by permuting pairs of atoms is a strict subset of $S_6$, which
    we denote by $S_4^{\rm D}$.}
   \label{fig:perm_group}
\end{figure}

For the $n$-body terms with $n \geq 4$, the permutation group $S_n$ acting on
$\{ \bR_i \}_{i = 1}^n$ which encodes the symmetry of $\E_n$ induces a non-trivial
permutation group $S_n^{\rm D} \subsetneq S_{n(n-1)/2}$ acting on
$\{\x_{ij}\}_{i < j = 1}^n$ encoding the symmetry of $\Er_n^{\rm D}$. That is,
a permutation $\pi \in S_n$ of sites $\{\bR_i\}$ is re-interpreted as a
permutation in $S_n^{\rm D}$ of distances through the action
\[
  \x_{ij} \overset{\pi}{\mapsto}  \x_{\pi i \pi j}
\]
see Figure~\ref{fig:perm_group} for a visualisation in the
four-body case.

Employing invariant theory techniques~\cite{Derksen2015-km,Braams2009-wi} we
then construct {\em fundamental invariants}~$f_n = \{ \If_{n,a} \}_{a = 1}^{A_n}$,
where $A_n > n$ is the dimensionality of the set $f_n$ and each $\If_{n,a}$ is a
multi-variate polynomial in $\{u_{ij}\}$ that is invariant
under $S_n^{\rm D}$, such that we can rewrite $\E_n$ as
\[
   \E_n(\{\bR_i\}_{i=1}^n) = \Er_n^{\rm D}(\{u_{ij}\})
      = \Erp_n^{\rm D}(\{ \If_{n,a} \}).
\]
A subtle point is that admissible arguments $\{u_{ij}\}$ and $\{\If_{n,a}\}$
belong to $(3n-6)$-dimensional submanifolds of, respectively, $\R^{n(n-1)/2}$
and $\R^{A_n}$, where $n(n-1)/2 > (3n-6)$ for $n \geq 5$ and $A_n > n(n-1)/2$
for $n \geq 4$; cf. Table~\ref{tab:coord_dims}.
For illustration,  a possible choice of the fundamental invariants for a 4-body
distance-based potential are given in Table~\ref{tbl:L-4b-invariants}.

\begin{table}
   \begin{tabular}{r|ccccc}
      $n$      &  2  &  3  &  4  & 5 & 6    \\
      \hline
      \#coordinates &  1  &  3  &  6  &  9 & 12  \\
     dim $\{u_{ij}\}$ &  1  &  3  &  6  &  10 & 15 \\
    $A_n$ = dim $\{\If_{n,a} \}$ &  1  &  3  &  9 &  56 & - \\
   \end{tabular}
   \caption{
      \label{tab:coord_dims}
      Comparison of potential minimal number of coordinates, number of RI
      coordinates (distances), and number of RPI coordinates (fundamental
      invariants). For six-body invariants {\sc Magma} terminated without computing the fundamental invariants within several weeks of CPU time. This is related to the rapidly increasing cardinality of the symmetry group (6! = 720). Note that restricting the polynomial degree shortens the computations, hence allows to compute invariants for higher body-orders, as is performed in~\cite{Qu2013-on}.
      }
\end{table}

\begin{table}
{\scriptsize
\begin{tabular}{rl}
   $ \If_{4,1} = $ & $ u_{12} + u_{13} + u_{14} + u_{23} + u_{24} + u_{34}$ \\
   $ \If_{4,2}= $ & $ u_{12}^2 + u_{13}^2 + u_{14}^2 + u_{23}^2 + u_{24}^2 + u_{34}^2 $ \\
   $\If_{4,3}= $ & $ u_{12}u_{13} + u_{12}u_{14}
   + u_{12}u_{24} + u_{12}u_{34}$ \\
   $   $ & $ + u_{13}u_{14} + u_{13}u_{23}  + u_{13}u_{34} + u_{14}u_{23} $ \\
   $ $ & $ + u_{14}u_{24} + u_{23}u_{24} + u_{23}u_{34} + u_{24}u_{34} $
    \\
    $
   \If_{4,4}= $ & $ u_{12}^3 + u_{13}^3 + u_{14}^3 + u_{23}^3 + u_{24}^3 + u_{34}^3
   $ \\ $
   \If_{4,5}= $ & $ u_{12}u_{13}u_{14} + u_{12}u_{24}u_{34} + u_{13}u_{23}u_{34} + u_{14}u_{23}u_{24}
   $ \\ $
   \If_{4,6}= $ & $ u_{12}^4 + u_{13}^4 + u_{14}^4 + u_{23}^4 + u_{24}^4 + u_{34}^4
   $ \\ $
   \If_{4,7}= $ & $ u_{12}^2u_{13} + u_{12}^2u_{14} + u_{12}^2u_{24} + u_{12}^2u_{34} + u_{12}u_{13}^2 $ \\ $  $ & $
   + u_{12}u_{14}^2
    + u_{12}u_{24}^2 + u_{12}u_{34}^2 +
    u_{13}^2u_{14} + u_{13}^2u_{23}
    $ \\ $
     $ & $ + u_{13}^2u_{34} + u_{13}u_{14}^2 + u_{13}u_{23}^2 + u_{13}u_{34}^2 + u_{14}^2u_{23} $ \\ $  $ & $
    +
    u_{14}^2u_{24} + u_{14}u_{23}^2 + u_{14}u_{24}^2 + u_{23}^2u_{24} + u_{23}^2u_{34} $ \\ $  $ & $
    + u_{23}u_{24}^2 + u_{23}u_{34}^2 +
    u_{24}^2u_{34} + u_{24}u_{34}^2
    $ \\ $
   \If_{4,8}= $ & $ u_{12}^2u_{13}u_{23} + u_{12}^2u_{14}u_{23} + u_{12}^2u_{23}u_{24} + u_{12}^2u_{23}u_{34} $ \\ $  $ & $
   + u_{12}u_{13}^2u_{24} + u_{12}u_{13}u_{23}^2 +
    u_{12}u_{13}u_{24}^2 + u_{12}u_{14}^2u_{34} $ \\ $  $ & $
    + u_{12}u_{14}u_{23}^2 + u_{12}u_{14}u_{34}^2 + u_{12}u_{23}^2u_{24} + u_{12}u_{23}^2u_{34} $ \\ $  $ & $
     + u_{13}^2u_{14}u_{24} + u_{13}^2u_{23}u_{24} + u_{13}^2u_{24}u_{34} + u_{13}u_{14}^2u_{34} $ \\ $  $ & $
     + u_{13}u_{14}u_{24}^2 +
    u_{13}u_{14}u_{34}^2 + u_{13}u_{23}u_{24}^2 + u_{13}u_{24}^2u_{34} $ \\ $  $ & $ + u_{14}^2u_{23}u_{34} + u_{14}^2u_{24}u_{34} + u_{14}u_{23}u_{34}^2  + u_{14}u_{24}u_{34}^2
   $ \\ $
   \If_{4,9}= $ & $ u_{12}^2u_{13}u_{23}^2 + u_{12}^2u_{14}u_{23}^2 + u_{12}^2u_{23}^2u_{24} + u_{12}^2u_{23}^2u_{34} $ \\ $  $ & $ + u_{12}u_{13}^2u_{24}^2 + u_{12}u_{14}^2u_{34}^2 + u_{13}^2u_{14}u_{24}^2 + u_{13}^2u_{23}u_{24}^2 $
   \\
   $  $ & $ + u_{13}^2u_{24}^2u_{34} + u_{13}u_{14}^2u_{34}^2 + u_{14}^2u_{23}u_{34}^2 + u_{14}^2u_{24}r_{34}^2. $
\end{tabular}}
\caption{\label{tbl:L-4b-invariants} 4B Distance-based invariants.}
\end{table}

To complete the definition of $n$-body distance-based potentials, we supply
$E_n$ with a cut-off mechanism, redefining them as
\begin{equation}
   \label{eq:fcut_BL}
   \E_n(\{\bR_i\}_{i=1}^n) = \Erp_n^{\rm D}(\If_n) \prod_{i < j} f_{\rm cut}(r_{ij}),
\end{equation}
where $f_{\rm cut}(r)$ is smooth and vanishes outside some cut-off radius
$r_{\rm cut}$. That is, we only account for $n$-body clusters for which
{\em all} edge lengths are within the cut-off radius.
The choice of cut-off mechanism is far from unique of course,
and can be adapted to the systems of interest.

\subsection{Distance-angle potentials}
\label{sec:bapots}
While distance-based coordinates are seemingly canonical in that they inherit
the maximum symmetry, it is sometimes intuitive to employ a coordinate system
that incorporates more ``physically natural'' coordinates, which may lead to
alternative $\RI$ and $\RPI$ coordinate systems. For example, the success of
employing bond-angle coordinates in empirical force
fields~\cite{Stillinger1985-ul} suggests that using these coordinates may
produce better fits at similar cost.  This idea is further supported by the
success of Moment Tensor Potentials~\cite{Shapeev2016-hn}, which can be also
interpreted as distance-angle potentials. Within our framework, many-body
distance-angle potentials are constructed as follows.

As in \S~\ref{sec:pots:BL} let $\x_{ij}$ denote transformed distance
coordinates. In addition, let $w_{ijk}$ denote angle coordinates, the
canonical choice being
\[
w_{ijk} = \cos\theta_{ijk} = \hat{\bR}_{ij} \cdot \hat{\bR}_{ik},
\]
where $\bR_{ij} = \bR_j -
\bR_i$ and $\hat{\bR} = \bR/|\bR|$. We then write an $n$-body term $E_n$ in
a way that retains only partial symmetry,
\[
  \E_n(\{\bR_i\}_{i=1}^n)
  = \Er_n^{\rm DA}( \{\x_{1j}\}_{j=2}^n,
  \{w_{1jk}\}_{j<k=2}^n \big),
\]
where the superscript ``DA'' indicates that the function $\Er_n^{\rm DA}$ is
parametrised by distances and angles. The 2-body contribution $\E_2$ is again a
pure distance-based potential.

For body order $n \geq 3$, we transform again to an RPI coordinate
system. For  the distance-angle potentials we retain only permutation invariance
with respect to the $n-1$ neighbours, which induces a different symmetry group
$S_{n-1}^{\rm DA}$ on the coordinates $(\{\x_{1j}\}_j, \{w_{1jk}\}_{jk})$.
Analogously to
the distance based potentials, the fundamental polynomial invariants
for this permutation group $S_{n-1}^{\rm DA}$ yield a RPI coordinate
system $\If_n = \If_n(\{\x_{1j}\}_j, \{w_{1jk}\}_{jk})$ and thus the
representation
\begin{align*}
   \E_n(\{\bR_i\}_{i=1}^n)
   =
   \Erp_{n}^{\rm DA}\big(\If_n\big).
\end{align*}
For 3-body potentials a possible choice of fundamental invariants is
\begin{align}
   \If_{3,1}(u_{12}, u_{13}, w_{123}) & = u_{12}+u_{13}, \nonumber \\
   \If_{3,2}(u_{12}, u_{13}, w_{123}) & = u_{12} u_{13},\nonumber \\
   \If_{3,3}(u_{12}, u_{13}, w_{123}) & = w_{213}, \label{eq:I3_BA}
\end{align}
while a possible set for four-body potentials is given in
Table~\ref{tbl:LA-4b-invariants}.

\begin{table}
{\scriptsize
\begin{tabular}{rl}
   $ \If_{4,1}=$ & $ u_{12} + u_{13} + u_{23}   $
   \\
   $ \If_{4,2}=$ & $ w_{213} + w_{214} + w_{314}   $
   \\
   $ \If_{4,3}=$ & $ u_{12}^2 + u_{13}^2 + u_{23}^2   $
   \\
   $ \If_{4,4}=$ & $ u_{12}w_{213} + u_{13}w_{314} + u_{23}w_{214}   $
   \\
   $ \If_{4,5}=$ & $ w_{213}^3 + w_{214}^3 + w_{314}^3   $
   \\
   $ \If_{4,6}=$ & $ u_{12}^3 + u_{13}^3 + u_{23}^3 + w_{213}^2w_{214} + w_{213}w_{314}^2 +  w_{214}^2w_{314}  $
   \\
   $ \If_{4,7}=$ & $ u_{12}w_{214} + u_{13}w_{213} + u_{23}w_{314}   $
   \\
   $ \If_{4,8}=$ & $ w_{213}^2 + w_{214}^2 + w_{314}^2   $
   \\
   $ \If_{4,9}=$ & $ u_{12}^2u_{23} + u_{12}u_{13}^2 + u_{13}u_{23}^2   $
   \\
   $ \If_{4,10}=$ & $ u_{12}u_{13}w_{213} + u_{12}u_{23}w_{214} + u_{13}u_{23}w_{314}   $
   \\
   $ \If_{4,11}=$ & $ u_{12}w_{213}^2 + u_{13}w_{314}^2 + u_{23}w_{214}^2   $
   \\
   $ \If_{4,12}=$ & $ u_{12}^2w_{214} + u_{13}^2w_{213} + u_{23}^2w_{314}   $
   \\
   $ \If_{4,13}=$ & $ u_{12}w_{213}w_{214} + u_{13}w_{213}w_{314} + u_{23}w_{214}w_{314}   $
   \\
   $ \If_{4,14}=$ & $ w_{213}^2w_{314} + w_{213}w_{214}^2 + w_{214}w_{314}^2.   $
\end{tabular}
}
\caption{\label{tbl:LA-4b-invariants} 4B Distance-angle-based invariants.}

\end{table}

The number of fundamental invariants is higher than for the
distance based coordinate system, while the degrees of the invariants are
smaller. Thus, the invariants are cheaper to compute but more basis
functions will be required.  This is due to the fact that we exploited less symmetry in the distance-angle potentials than in the purely
distance based potentials.

Finally, we propose a natural cut-off mechanism for distance-angle potentials,
\begin{equation}
   \label{eq:fcut_BA}
   \E_n(\{\bR_i\}_{i=1}^n)
   =
   \Erp_{n}^{\rm DA}\big(\If_n\big)
   \prod_{j = 2}^n f_{\rm cut}(r_{1j}).
\end{equation}
This cut-off mechanism is different from the one in the distance-based potential
\eqref{eq:fcut_BL} since it
only acts on the distance
variables and thus the product is taken over a smaller set of $r_{ij}$
values, leading to a summation over a different set of $n$-body clusters
in the total potential energy assembly. In particular, the meaning of the cutoff
radius of $f_{\rm cut}$ in \eqref{eq:fcut_BL} and \eqref{eq:fcut_BA} is not
equivalent.

\subsection{Polynomial Approximation}
In the following, we write $\Erp_n$ to mean $\Erp_n^{\rm D}$ or $\Erp_n^{\rm DA}$ depending
on whether we are considering distance or distance-angle coordinates.
To construct computable representations of the $E_n^\RPI$ we will use
multi-variate polynomials: the $n$-body functions $E_n$ will be
represented as polynomials of the invariants $f_n = \{f_{n,j}\}_{j=1}^{A_n}$, i.e.,
\[
  E_n(\{\bR_i\})
  =
  \Erp_n(\If_n)
  =
  P_n(\If_n),
\]
where $P_n$ is a multi-variate polynomial in $f_n$ with coefficients
that are to be determined; see \S~\ref{sec:lsq}.
  By increasing the polynomial degree of $P_n$ we
can in principle approximate arbitrary smooth, symmetric functions
$E_n$. In particular, letting the body-order $n$, the
cutoff $r_{\rm cut}$ and the polynomial degrees all tend to
infinity we can represent an arbitrary smooth PES. That is, our
construction is {\em systematically improvable}.

However, we briefly discuss a subtlety that arises for $n \geq 4$ when the
permutation groups become non-trivial: in this case, the representation of a
given symmetric polynomial in terms of the fundamental invariants is not unique.
This non-uniqueness can be avoided by introducing an alternative set of
invariants, the \textit{primary invariants} $\Ip_n = \{\Ip_{n,a}\}_{a = 1}^{n(n-1)/2}$ and \textit{secondary
invariants} $\Is_n = \{\Is_{n,b}\}_b$ both of which can be constructed from the fundamental invariants~\cite{Derksen2015-km,Braams2009-wi}. In terms of $\Ip_n, \Is_n$
\begin{equation}
   \begin{split}
   \label{eq:RPI_Prim_Sec}
   \E_n(\{\bR_i\}_{i=1}^n)
   &= P_n(\If_n)   \\
   &= \sum_b \Is_{n,b} P_{n,b}(\Ip_n),
   \end{split}
\end{equation}
where each $P_{n,b}$ is a multivariate polynomial in $n(n-1)/2$ variables of the set
 $\Ip_{n}$,
and the summation ranges over all secondary invariants.
Once the invariant sets $p_n, s_n$
are specified, this decomposition of a symmetric polynomial is unique,
which gives a simple way to generate all symmetric polynomials with a
prescribed degree. The choice of invariants $p_n, s_n$ remains
non-unique.  A ``manual'' construction for $n=4$ is proposed
by~\citet{Schmelzer1985-fe}, however, for $n > 4$ this can practically only be
achieved using a computer algebra system; we employ the {\sc Magma}~software
package~\cite{Bosma1997-pq}.

We briefly describe the primary and secondary invariants for 3- and
4-body terms. For 3-body potentials, the permutation group is trivial
(all of $S_3$), hence the $\Ip_3 = \If_3$ and $\Is_3 = (1)$. For
4-body potentials, the primary invariants  are
\[
  \Ip_4 =
  \{ \If_{4,a}\}_{a=1}^6,
\]
  which thus depend on the choice of
coordinate system, D or DA, through the definition of $f_n$.
The secondary invariants
also depend on the choice of RI coordinates. For distance-based potentials
there are six secondary invariants,
\[
   \Is_4 = \big(1, \If_{4,7}, \If_{4,8},
   \If_{4,9}, \If_{4,7}^2, \If_{4,8} \If_{4,9}\big),
\]
while for distance-angle potentials there are twelve secondary invariants,
\begin{align*}
   \Is_4 =  \big( & 1, \If_{4,7}, \If_{4,8}, \dots, \If_{4,14},
      \If_{4,7} \If_{4,8}, \If_{4,8}^2, \If_{4,10}^2 \big)
\end{align*}
Note that the constant polynomial $1$ is usually not considered as a secondary invariant;
we include it for notational convenience.

For five-body potentials in distance based coordinates, we find 144 secondary
invariants out of which 21 are irreducible. For distance-angle potentials there
are 266 secondary invariants among which 44 are irreducible. Due to the
complexity and large numbers of these polynomials we use the {\sc Magma} output
to {\em auto-generate} source-code that evaluates the invariants and their
derivatives for our aPIP implementation.

\subsection{Least-squares fit}
\label{sec:lsq}
All variants of the body-ordered interatomic potentials
$\E(\RR)$ we proposed in the foregoing sections can be
expressed as a linear combination of basis functions $B_{nb{\bf k}}$,
where $n$ is the body-order, ${\bf k} \in \N^{n(n-1)/2}$ defines a
monomial in the primary invariants and $b\in\N$ denotes the indices of
the secondary invariant in \eqref{eq:RPI_Prim_Sec}.

To see this, we first specify a total polynomial degree $D_n > 0$, and recall
that the invariants $p_{n,a}$ and $s_{n,b}$ are themselves polynomials in the RI
coordinates and have therefore a well-defined total degree ${\rm deg}(p_{n,a}),
{\rm deg}(s_{n,b})$. We can now write the polynomials  $P_{n,b}$ from
\eqref{eq:RPI_Prim_Sec} as
\[
  P_{n,b}(\{p_{n,a}\})
  =
  \sum_{\{k_a\}}
  c_{nb{\bf k}}
  \prod_{a = 1}^{n(n-1)/2} (p_{n,a})^{k_a},
\]
where the summation
ranges over all tuples ${{\bf k}} = (k_a)_{a = 1}^{n(n-1)/2}$ of non-negative
integers with
\[
  {\rm deg}(s_{n,b}) + \sum_a k_a {\rm deg}(p_{n,a}) \leq D_{n}.
\]
The coefficients $c_{nb{\bf k}}$ are the unknowns to be determined in the
least squares fit. Thus we see that for each
body-order $n$, each secondary invariant indexed by $b$ and for each
tuple ${\bf k}$ specifying a monomial within the prescribed degree $D_n$
we obtain a corresponding basis function for the total potential energy
\begin{align}
   B_{nb{\bf k}}(\{\bR_i\}_{i=1}^M)
   & =  \sum_{\substack{1 < i_1 < i_2 \\ < \ldots < i_n \le M}} F_{\rm cut}(\{\bR_{i_l}\}_{l=1,\ldots,n}) \\
   \notag
   & \qquad \qquad\times \left[ \Is_{n,b} \prod_{a=1}^{n(n-1)/2}  (\Ip_{n,a})^{k_a}  \right],
\end{align}
where $\Is_{n,b}, \Ip_{n,a}$ are evaluated at $\{\bR_{i_l}\}_{l=1,\ldots,n}$ and
the definition of the cut-off function $F_{\rm
  cut}(\{\bR_{i_l}\}_{l=1,\ldots,n})$ depends on the choice of
variables and is defined in \eqref{eq:fcut_BL} for the distance-based
case and \eqref{eq:fcut_BA} for the distance-angle case.  In the summation,
only clusters respecting the cut-off condition $F_{\rm
  cut}(\{\bR_{i_l}\}_{l=1,\ldots,n})>0$ taken into account to ensure
linear scaling cost.

It remains to determine the coefficients $c_{nb {\bf k}}$ in the linear expansion
\begin{equation}\label{eq:aPIP_PES_basis_expansion}
   \E(\{\bR_i\}_{i=1}^M) = \sum_{n, b, {\bf k}}
      c_{nb{\bf k}} B_{nb{\bf k}}(\{\bR_i\}_{i=1}^M),
\end{equation}
achieved via solving a linear least squares problem.

  For each atomic configuration
${\bf R}$ in a training set $\mathcal{R}$, the corresponding energy $\Ee_{\bf R}$, forces $\Fe_{\bf R}$ and possibly
virials $\Ve_{\bf R}$ are given. The minimized functional is of the form
\begin{equation} \label{eq:lsq_fcnl}
\begin{split}
   J =
   \sum_{{\bf R} \in \mathcal{R}} \Big( &\;
      W_{E}^2 \left| \E({\bf R}) - \Ee_{\bf R} \right|^2 \\[-2mm]
      & \;
      + W_{F}^2 \left| F({\bf R}) - \Fe_{\bf R} \right|^2 \\
      & \;
      + W_{V}^2 \left| V({\bf R}) - \Ve_{\bf R} \right|^2\Big),
\end{split}
\end{equation}
where $W_E$, $W_F$, $W_V$ are weights that may depend on the configurations
${\bf R}$, and $F({\bf R})$ and $V({\bf R})$ are respectively forces and virials
computed from the energy functional $\E({\bf R})$.

In summary, since $J$ is quadratic in the unknown polynomial coefficients
${\bf c} = \{c_{nb{\bf k}}\}$, its
minimisation is a standard linear least-squares problem
\begin{equation}
   \label{eq:LSQ_system}
   \min_{\bf c} \|A{\bf c} - Y\|_2^2,
\end{equation}
which we solve using a QR factorisation. The size of the system matrix
$A$ is $N_{\rm obs} \times N_{\rm basis}$ where $N_{\rm basis}$ is the
number of basis functions while $N_{\rm obs}$ is the number of
observations (energies, forces, virials, and regularisation, if any,
see below). In our examples $N_{\rm obs}$ may be in the range of
hundreds of thousands, however, $N_{\rm basis}$ remains low; on the
order of hundreds to a few thousands. In this case, the QR
factorisation of the matrix $A$ is computationally cheap ($O(N_{\rm
  obs} \times N_{\rm basis}^2)$ operations) and numerical stable. We
postpone discussion of regularisation mechanisms to the next section,
some of which will show up as a regularisation functional added to
$J$, which will thus remain a quadratic functional in ${\bf c}$.

\subsection{Systematic Convergence}
\label{sec:convergence_rmse}
The prospective accuracy of an interatomic potential is directly
related to its functional form, in our case the choice of basis
functions to represent the PES. The family of potentials we
constructed in the previous sections are systematically improvable: by
increasing the body-order, cutoff radius and polynomial degree they
are {\it in principle} capable of representing an arbitrary many-body
PES to within arbitrary accuracy. To support this claim, we begin by
studying the convergence of the root mean square error (RMSE) on two previously published training sets for tungsten and silicon.

We measure the convergence of the RMSE against two key features:
Firstly, the number of basis functions used to construct the potential
gives a crude measure of the cost of the training. Secondly, we
compare the accuracy of the fit against the evaluation time of the
forces, that is, the cost of one molecular dynamics step.

For both training sets, we demonstrate the convergence of the
potential for both the distance-based and the distance-angle
descriptors. For all potentials, the distance transform used is a
polynomial transform, that is $u_{ij} = (r_{\rm nn} / r_{ij})^{p}$
where $r_{\rm nn}$ is an estimate for the nearest-neighbour distance
($r_{\rm nn} = 2.74$~\AA\ for W and $r_{\rm nn} = 2.35$~\AA\ for
Si) and $p$ may vary with the body-order. The cutoff function is given
by
\begin{equation} \label{eq:poly_cutoff}
   f_{\rm cut}(r) =
    \begin{cases}
        \left[\left(r / r_{\rm cut}\right)^2 - 1\right]^2, & 0 \leq r < r_{\rm cut}, \\
        0, & \text{otherwise.}
    \end{cases}
\end{equation}
where $r_{\rm cut}$ is a cut-off radius that may again vary with the body-order.
The parameters for the individual potentials and the least squares
regression weights are given in~the supplement~\cite{supp_mat}.

To choose the functional form of the distance transforms and cutoff function we
first performed low-accuracy fits that showed that the fit accuracy varies
little across different choices of cut-off function and distance transform; see
the supplement~\cite{supp_mat}. However, we will find that distance-angle potentials achieve a
higher accuracy at comparable computational cost than distance-based potentials,
both on the W and Si training sets.

\subsubsection{Results for Tungsten}
We now present convergence results for a tungsten training set used for a
previously published SOAP-GAP model~\cite{Szlachta2014-tb}, generated with
CASTEP~\cite{CASTEP2005}, that consists of 9693 configurations including
primitive unit cells, surfaces, $\gamma$-surfaces, vacancies and dislocation
quadrupoles. Every configuration provides one total energy and 3$N_{at}$
force components where $N_{at}$ is the number of atoms per configuration. Some
configurations also provide six virial components. The resulting total number of
scalars used for the fit was 497271.

\begin{figure*}[]
   \begin{subfigure}[b]{0.5\textwidth}
      \centering
      \includegraphics[width=8cm]{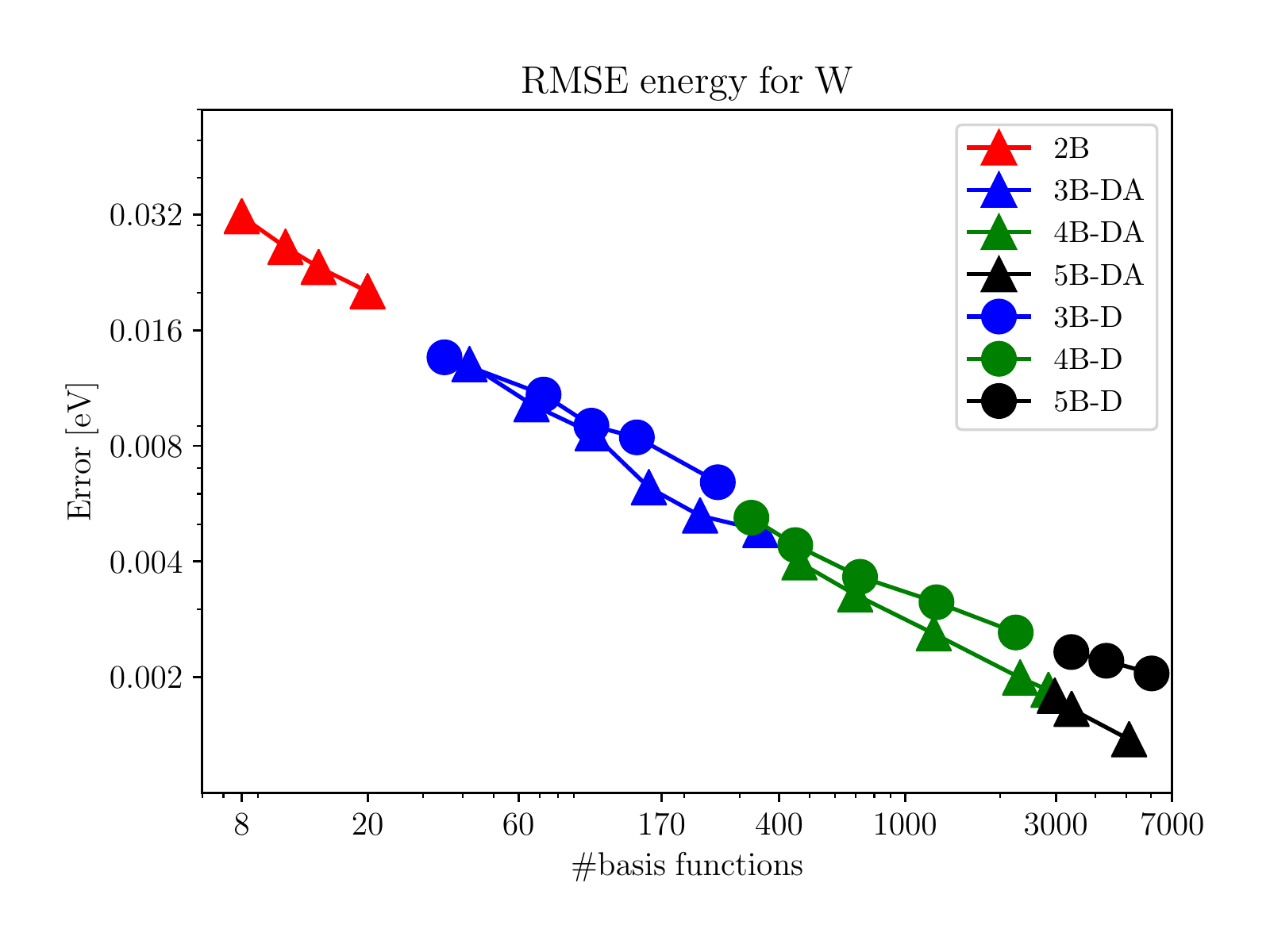}
      \caption{Convergence of the RMS error on the energy for a tungsten database with distance-angle descriptors (triangles) and distance-based descriptors (circles).
   }
   \label{fig:W_conv_error}
   \end{subfigure}%
   \begin{subfigure}[b]{0.5\textwidth}
      \centering
      \includegraphics[width=8cm]{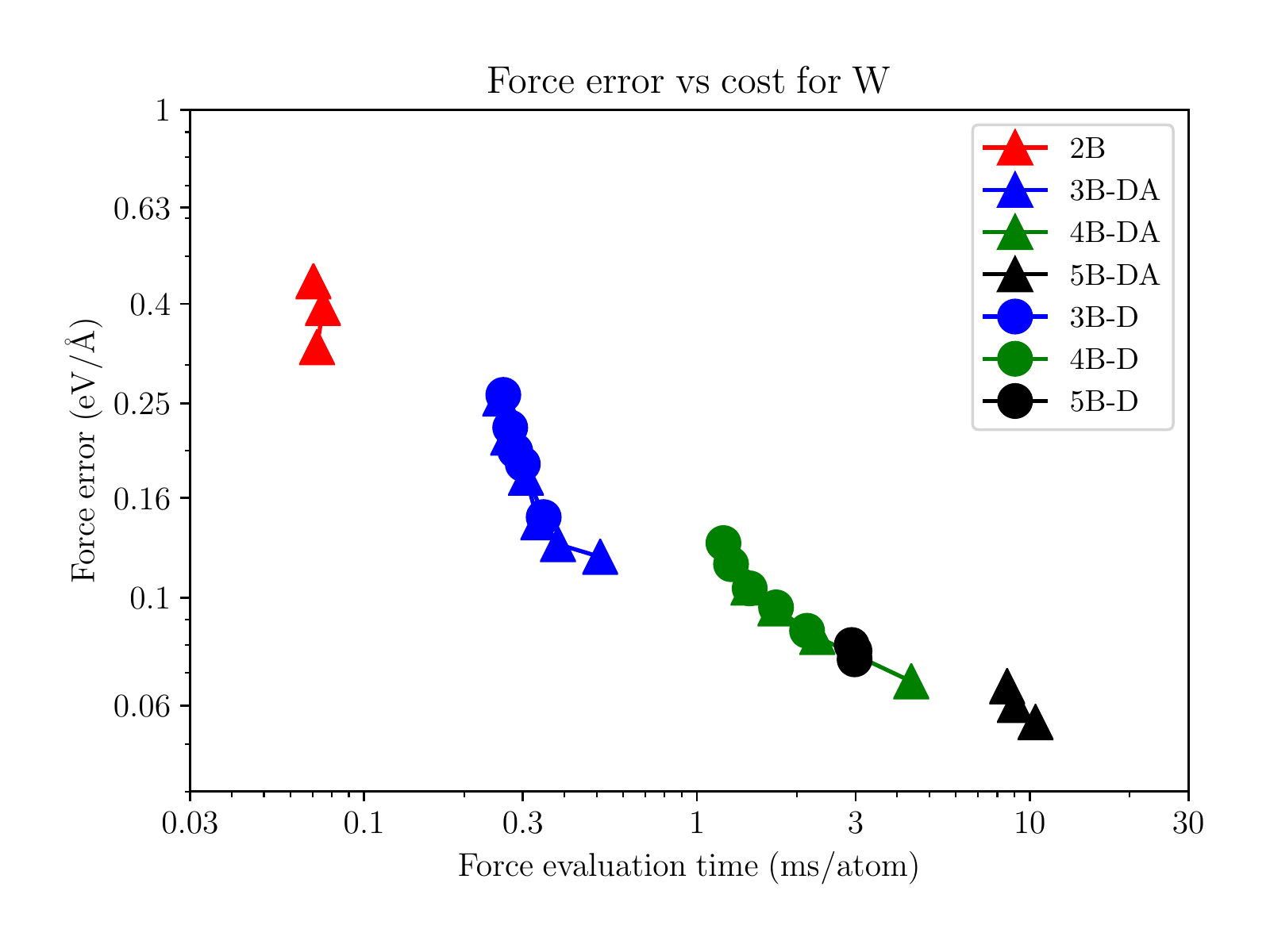}
      \caption{Timing for the force evaluation per atom for a tungsten database with distance-angle descriptors (triangles) and distance-based descriptors (circles).
      }
      \label{fig:W_conv_time}
   \end{subfigure}%

   \begin{subfigure}[b]{0.5\textwidth}
      \centering
      \includegraphics[width=8cm]{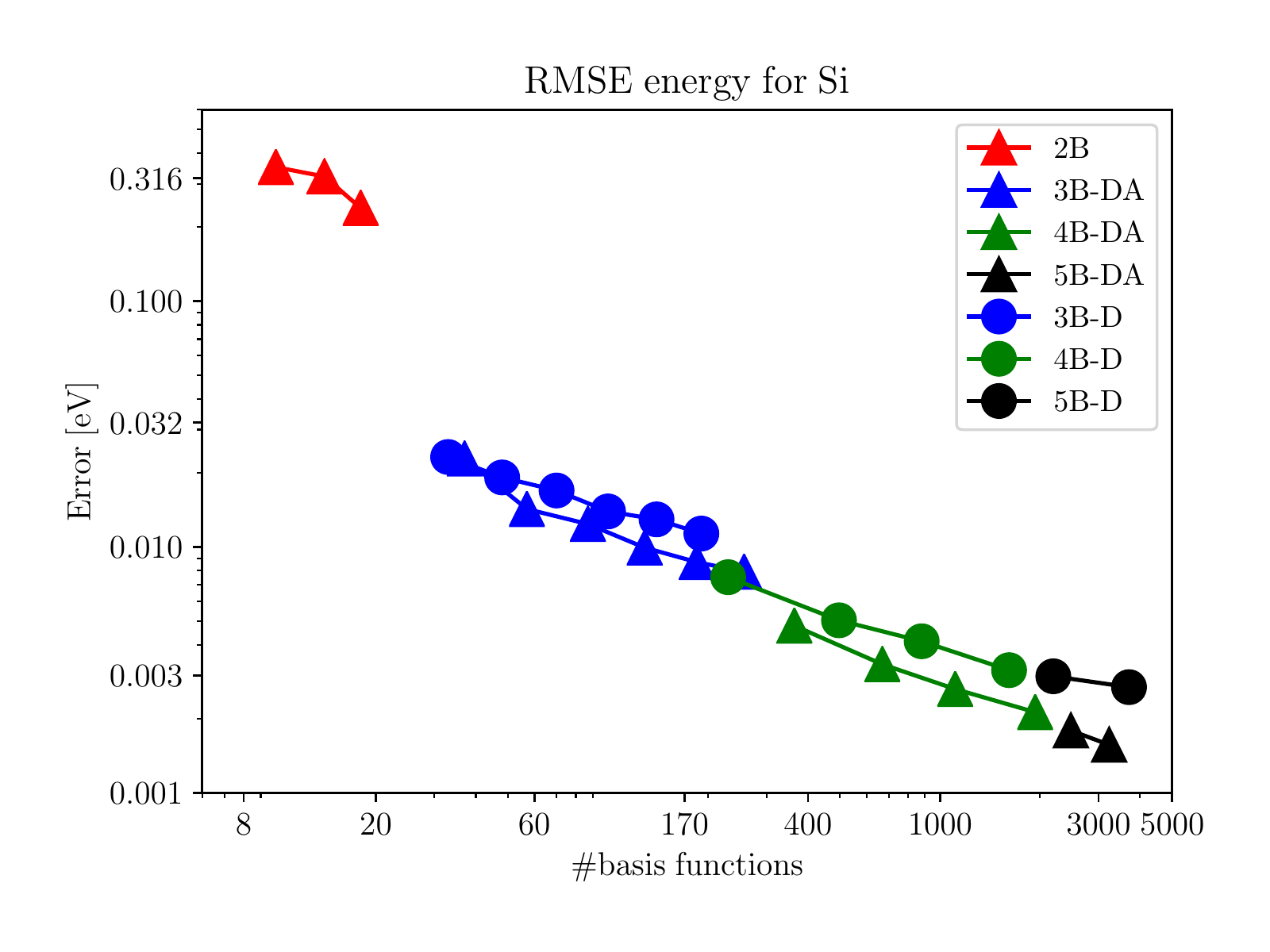}
   \caption{Convergence of the RMS error on the energy for a silicon database with distance-angle descriptors (triangles) and distance-based descriptors (circles).
    }
   \label{fig:Si_conv_error}
   \end{subfigure}%
   \begin{subfigure}[b]{0.5\textwidth}
      \centering
      \includegraphics[width=8cm]{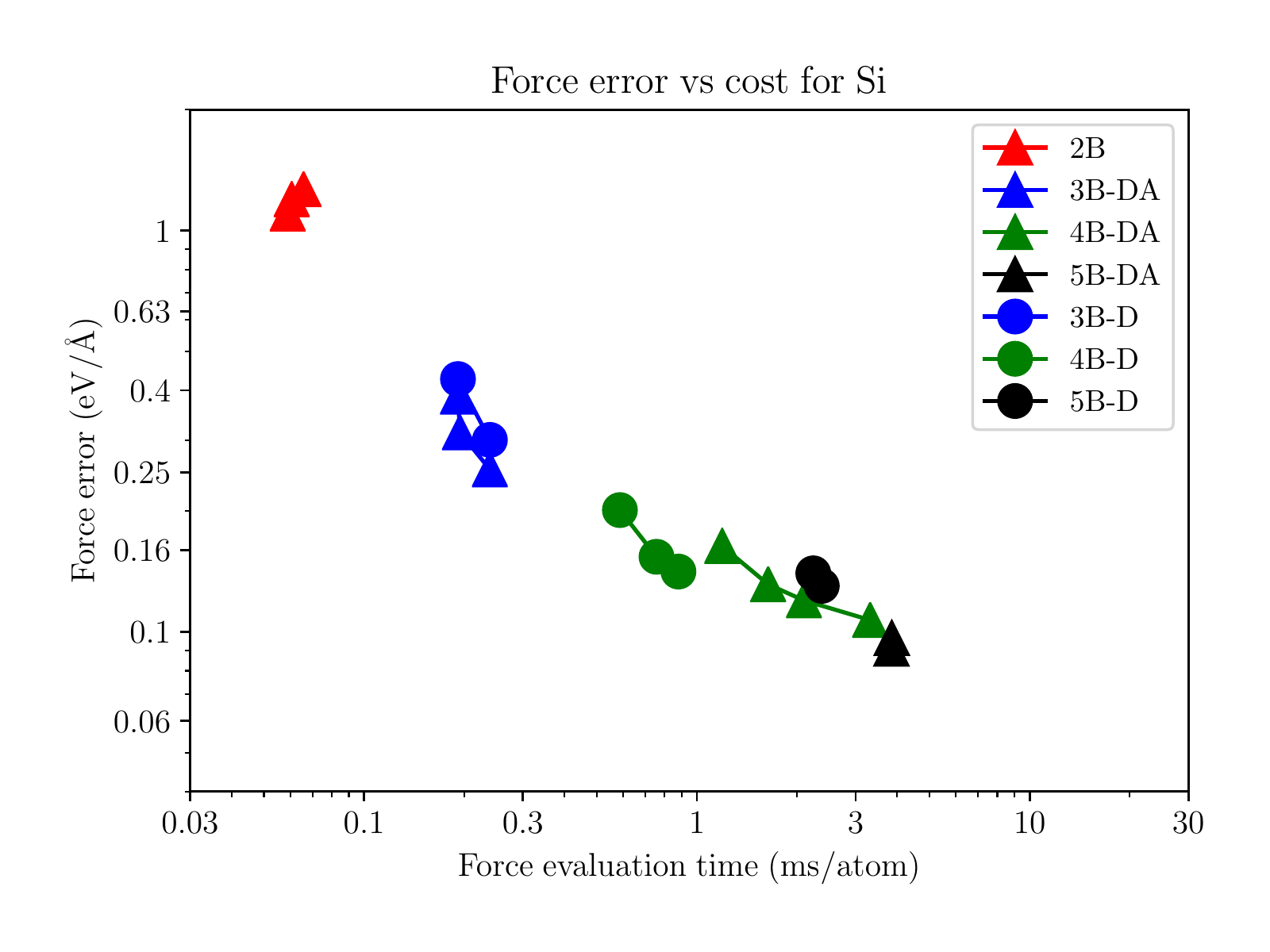}
   \caption{Timing for the force evaluation per atom for a Silicon database with distance-angle descriptors (triangles) and distance-based descriptors (circles).}
   \label{fig:Si_conv_time}
   \end{subfigure}%
      \caption{Systematic convergence on a tungsten (top) and a silicon (bottom) database.
          Note that the SOAP-GAP model takes about 30~ms / atom to evaluate on
a range of current CPUs, and MTP is 1~ms or faster depending on the number of basis functions used. }
\end{figure*}

For both distance-based and distance-angle potentials we observe in
Figure~\ref{fig:W_conv_error} the systematic decrease of the RMSE as the
body-order and the polynomial degrees are increased. Extended convergence tables
are presented in the supplementary material~\cite{supp_mat}.

In this test, distance-angle potentials perform slightly better
than distance-based potentials, particularly in the high accuracy
regime. Indeed, the distance-based potentials with 5-body reach an
energy RMSE of 2.05 meV with 6023 basis functions and 2.98~ms force
evaluation time per atom, while the distance-angle potentials for
4-body reach an energy RMSE of 1.85 meV with 2842 basis functions and
4.41~ms force evaluation time per atom. Thus, both the errors and
computational costs are comparable. The 5-body distance-angle
potentials reach an energy RMSE of 1.38 meV with 5113 basis functions
and 10.4~ms force evaluation time per atom.

\subsubsection{Results for Silicon}
We now demonstrate the convergence of the potential on a previously published silicon training set~\cite{Bartok:2018ih} which contains 2475 diverse configurations. We restrict the
published database to train only on the following subset of configurations: diamond cubic,
amorphous, $\beta$-tin, vacancies, sp2, and low index surfaces. The total amount of scalars included
in the fit (total energies, force/virial components) for this subset of the silicon database is 323414.
Although there are fewer configurations overall than in the tungsten database,
there are two distinct solid phases, and the amorphous phase
which is particularly challenging to fit. On the one hand, excluding certain parts of the published database
allows us to explore extrapolation. On the other hand, efficiently and accurately fitting to the complete
training set including a large variety of high coordination phases will likely require a more flexible
functional form, which we will revisit in future work.

The full convergence tables are presented in the supplementary material~\cite{supp_mat}.

As for tungsten, the
choice of descriptors gives similar accuracy and evaluation times for silicon,
but distance-angle potentials now reach significantly lower errors for large
basis sets. Moreover, the convergence plots presented on
Figures~\ref{fig:Si_conv_error} and~\ref{fig:Si_conv_time} show a systematic
convergence of the energy error for distance-angle and distance-based
descriptors. More precisely, the accuracy reaches the value of 2.13~meV accuracy
for a distance-angle 4-body potential composed of 1933 basis functions with a
force evaluation time of 3.32~ms per atom, and for a distance-based 5-body
potential, the energy error reaches 2.49~meV composed of 5396 basis functions
with a force evaluation time of 2.51~ms per atom. Finally, the 5-body
distance-angle potentials reach an energy error of 1.47~meV with 3759 basis
functions and a force evaluation time of 3.91~ms per atom.

\section{Regularisation}

\subsection{Regularisation Techniques}
\label{sec:regularisation}
In the least-squares method, regularisation is primarily seen as a
procedure to improve conditioning on ill-conditioned or even ill-posed
problems. By contrast, in the Gaussian process framework, it can be
interpreted as imposing `prior' information about the potential energy
surface, in particular its regularity. Robust heuristics for choosing
the strength of the regularisation were crucial for the success of the
GAP scheme for materials, where the regulariser was chosen to be
consistent with the estimated convergence error in the input data e.g. with
respect to k-point sampling~\cite{Dragoni:2018je}.

In the following we seek to apply a similar perspective in the standard least
squares framework. We will show how the low-dimensional functional
forms obtained in our definition of aPIPs in \S~\ref{sec:pots} allow
us to incorporate physically motivated `prior' information or
requirements that are not present in the database.

\begin{figure*}
   \begin{subfigure}[b]{0.32\textwidth}
      \centering
      \includegraphics[width=\textwidth]{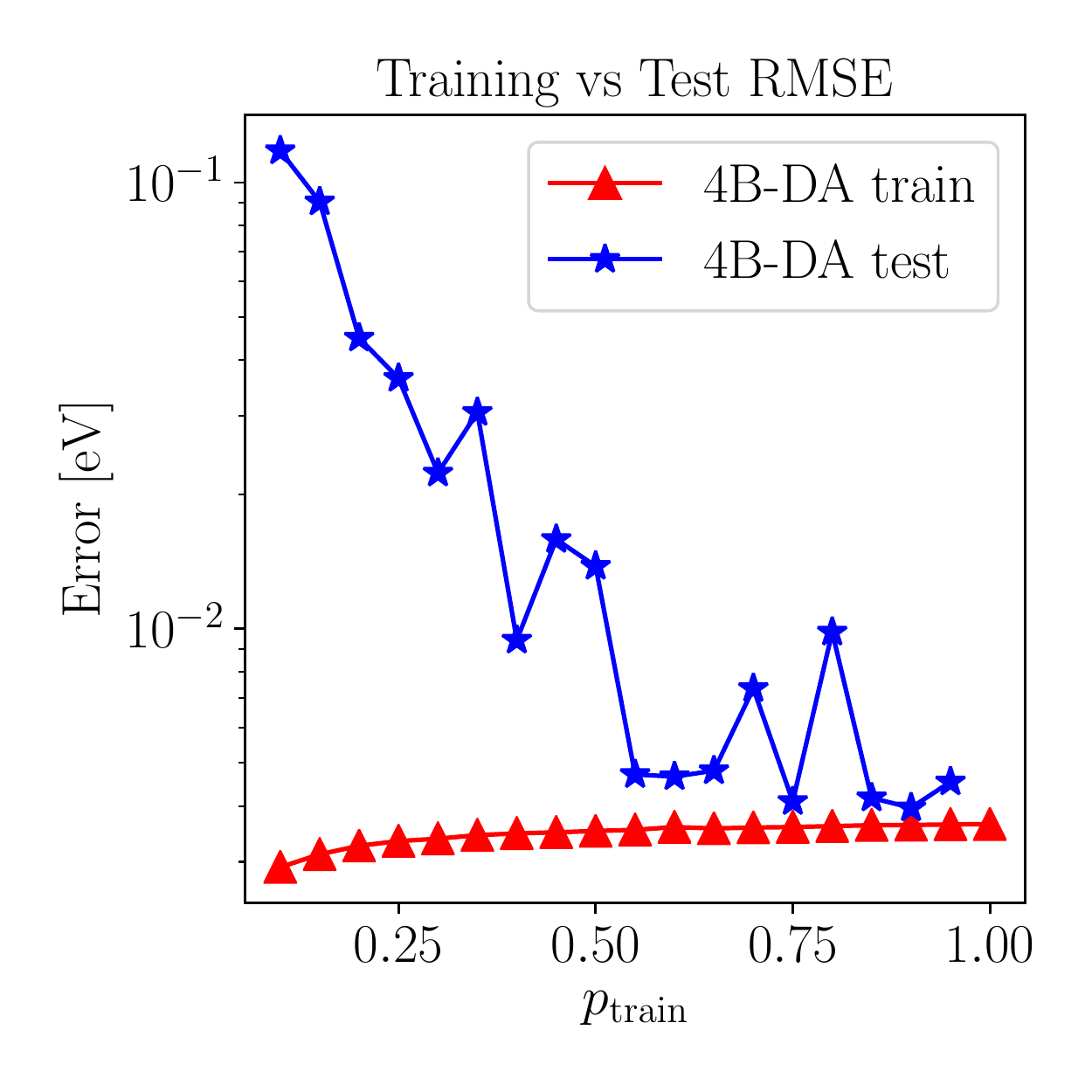}
      \caption{Training and test errors for a series of unregularised 4B-DA potentials for tungsten. Here, $p_{\rm train}$ is the proportion of training configurations, chosen at random.
 \\
      \quad \\
   }
   \quad
   \label{fig:train-test}
   \end{subfigure}
   \begin{subfigure}[b]{0.32\textwidth}
      \centering
      \includegraphics[width=\textwidth]{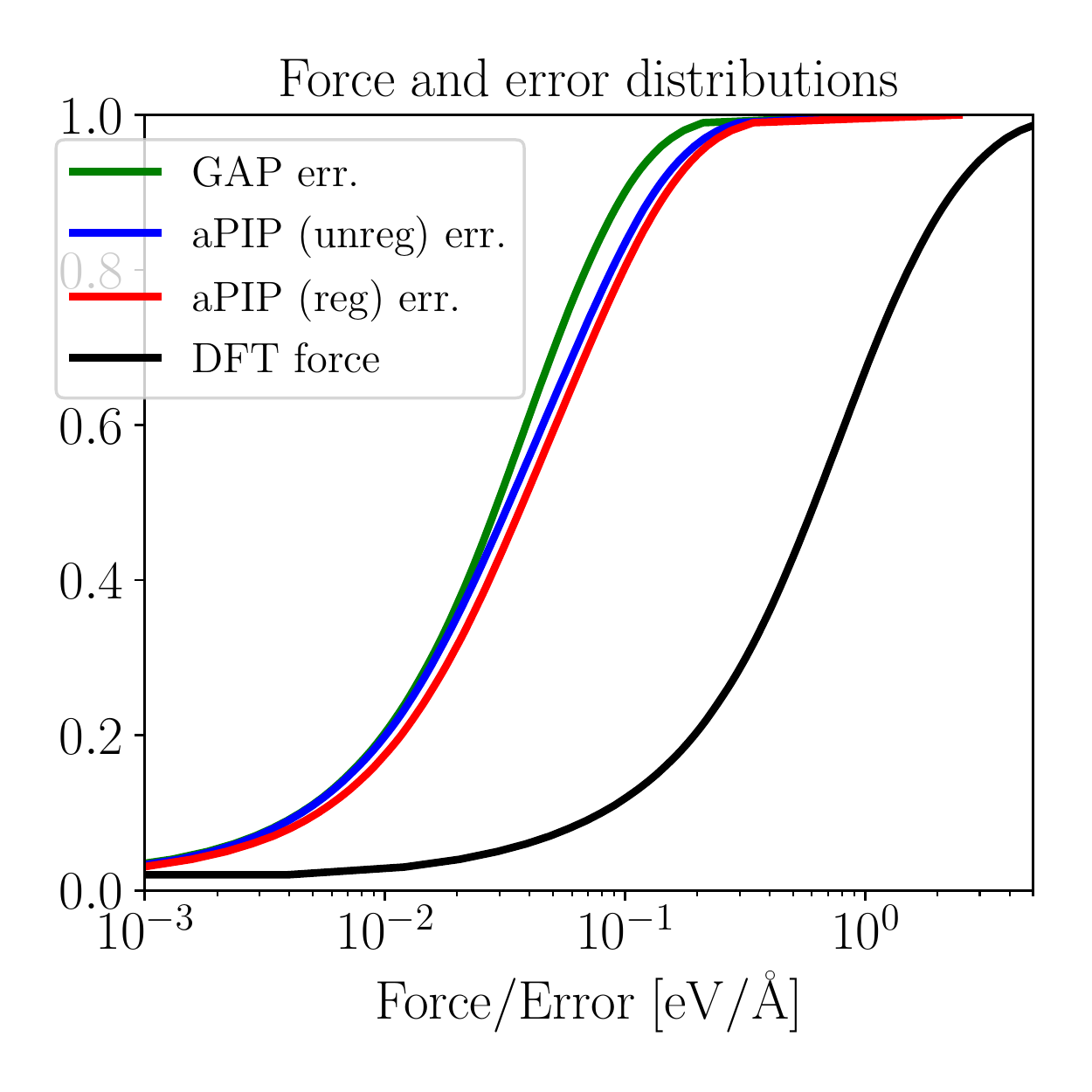}
      \caption{Cumulative force error distribution on the entire tungsten training set. The aPIP potentials here correspond to $p_{\rm train}=1$ of the other panels, and the SOAP-GAP model is the one originally published with the data set.\\
   }
   \label{fig:cumulative}
   \end{subfigure}
   \quad
   \begin{subfigure}[b]{0.32\textwidth}
      \centering
      \includegraphics[width=\textwidth]{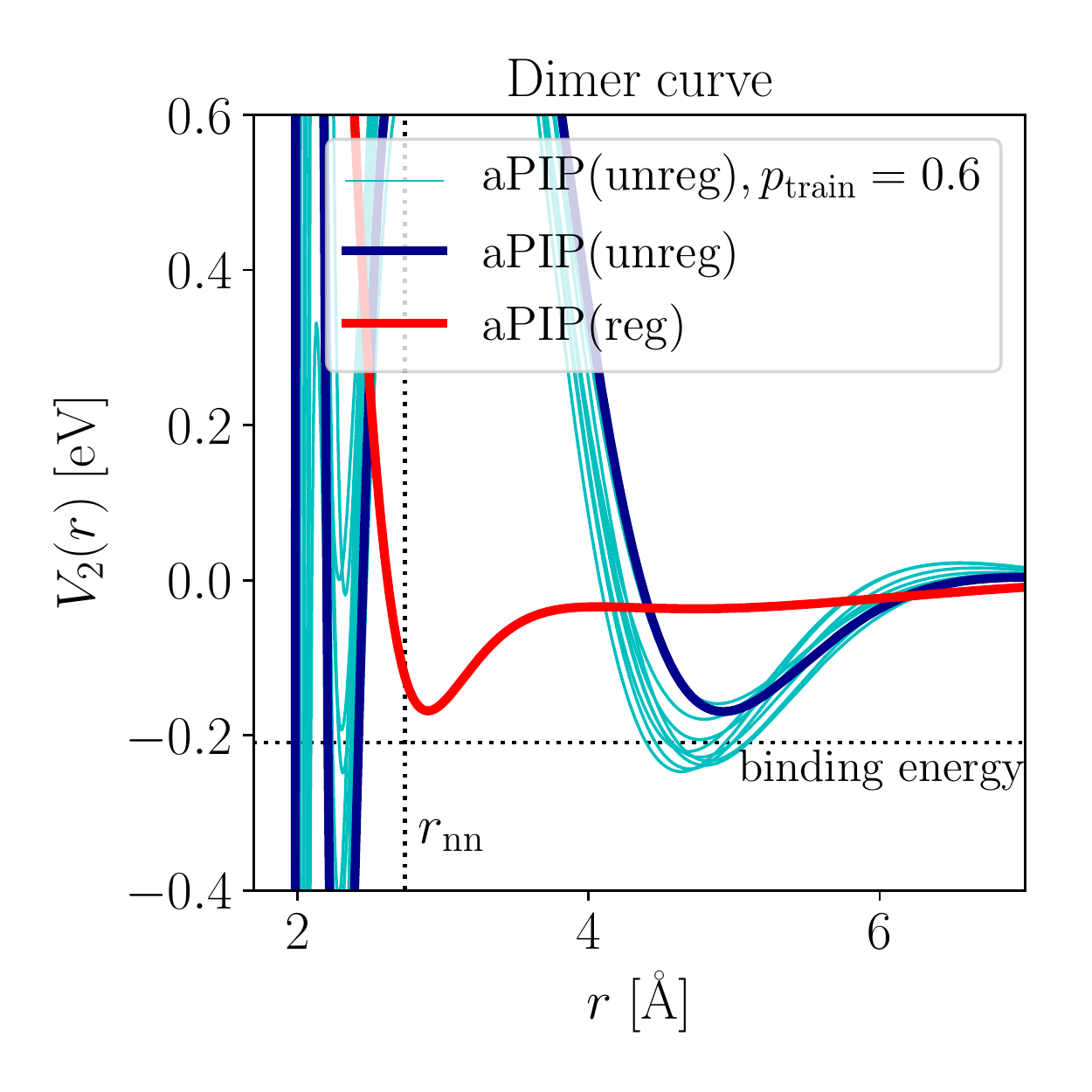}
      \caption{Two-body part of the unregularised aPIP potentials with $p_{\rm train} = 0.6$  (multiple curves correspond to different subsets) and $p_{\rm train} = 1.0$, as well as the regularised potential; see \S~\ref{sec:regularisation:fitaccuracy} for the details. \\
      \quad \\ }
   \label{fig:rough_vs_reg_2b2}
   \end{subfigure}%
   \label{fig:train-test-total}
   \caption{Comparison of training and test errors, and the corresponding two-body potentials.}
\end{figure*}

For example, consider the unregularised pair potential fit to the W database
displayed in Figure~\ref{fig:rough_vs_reg_2b2}, which is obtained by fitting a
degree 16 polynomial with distance transform $u_{ij} = (2.74 \text{\AA} /
r_{ij})^{2}$ and cutoff radius $r_{\rm cut} = 8.5 \text{\AA}$; see
\S~\ref{sec:regularisation:fitaccuracy} for more details. While it gives low
RMSE on our dataset, it is a nonsensical pair potential that is unsuitable for
materials modelling work.

A typical approach to detect overfitting and validate the generalisation
capabilities of the fitted potential is to first separate the data into a
training set and a test set, then to perform the regression using only the
training set, and finally compute the errors separately on the training and on
the test set. A transferable fit should have comparable training and test
errors.

While such a procedure helps to prevent overfitting {\em near} the
training set, we find that it gives very limited information about the ability
of the potential to generalise more broadly. While training and test errors are
comparable for a proportion of training configurations of 0.6 and higher (Figure
\ref{fig:train-test}), all potentials exhibit an oscillatory shape (Figure
\ref{fig:rough_vs_reg_2b2}) which is pathological and clearly does not allow for
extrapolation. In addition, the pair potential minimum is far from the
nearest-neighbour distance. Therefore, the generalisation tests presented in
Section \ref{sec:extrapolation} are performed directly on physical properties
and not using a training/test splits.

We now introduce a range of tools that enable us to produce
regularised aPIP fits that retain RMSE accuracy close to the
unregularised aPIPs, but become highly transferrable potentials that
``extrapolate well'' and in particular have no regions of ``holes'' as
those described in~\cite{Nandi2019-bl}.

\subsubsection{Tikhonov regularisation}
First, we recall some background on regularisation. In the context of linear
least squares, the problem \eqref{eq:LSQ_system} is replaced with the regularised least squares problem
\begin{equation} \label{eq:reg_lsq}
   \min_{{\bf c}} \|A {\bf c} - Y \|_2^2 + \| \Gamma {\bf c} \|_2^2,
\end{equation}
where $\Gamma$ is called the Tikhonov matrix. The form $\| \Gamma {\bf c}
\|_2^2$ may be used to represent any positive quadratic functional acting on the aPIP
potential energy surface $E$ given by \eqref{eq:aPIP_PES_basis_expansion}.
The most common choice is $\Gamma = \alpha I$ ($L^2$-regularisation) where the
unknown parameter $\alpha$ may be obtained through ad hoc procedures, or related to the uncertainty of the data via the
Bayesian interpretation of the least squares problem.

Such regularisation techniques are, for example, employed to render ill-posed
problems well posed, or improve the conditioning of severely ill-conditioned
problems.
 In our context, polynomial basis functions generally lead to
ill-conditioning, which is exacerbated by the fact that the space of $n$-body
functions contains $m$-body functions with $m <n$, leading to a
near-degeneracy that is only partly alleviated by using of a different cut-off
and distance transform at each body-order.

To solve the regularised least squares problem we re-interpret it as a
standard least squares problem through the equivalent formulation
\[
  \min_{{\bf c}} \left\| \left[ \begin{matrix} A \\ \Gamma \end{matrix} \right] {\bf c}
    - \left[ \begin{matrix} Y \\ 0 \end{matrix} \right] \right\|_2^2,
\]
which is then solved using the QR factorisation.

\subsubsection{Rank-revealing QR factorisation}
$L^2$-regularisation can be effectively replaced by the rank-revealing QR
factorisation (rr-QR)~\cite{Chan1987-gu}, a decomposition which reveals the
near-degeneracy of the matrix $A$. The factorisation reads
\[
   A P = Q \begin{pmatrix}
      R_{11} & R_{12} \\
      0 & R_{22}
   \end{pmatrix},
\]
where $P$ is a permutation matrix, $Q$ is an orthogonal matrix,
$R_{11}$ and $R_{22}$ are upper triangular matrices and, importantly,
a given norm of the matrix $R_{22}$ is below some prescribed
tolerance.

Truncating $R_{22}$ in the resolution of the least-square system can
be seen as a regularisation as it removes the small modes in the
matrix $A$. To demonstrate this, let us compare rr-QR and
$L^2$-regularisation on a simple example, where $A$ is the nearly
rank-deficient matrix
$$
A = \begin{pmatrix} 1 & 0 \\
0 & \varepsilon  \end{pmatrix}, \quad \varepsilon \mbox{ small,}
$$
and the observations are $Y = (y_1,y_2)^T$.
The solution of the unregularised least squares problem is $c_1 = y_1, c_2 = y_2/\varepsilon$.
The rr-QR algorithm with a parameter $\alpha$ will instead compute
$$
c_1 = y_1, \quad  c_2 = \varepsilon^{-1}\delta_{\varepsilon > \alpha}  y_2,
$$
while the least squares solution with $L^2$-regularisation with parameter $\alpha$ is given by
$$
c_1 = \frac{1}{1+\alpha^2} y_1, \quad  c_2 = \frac{\varepsilon}{\varepsilon^2 + \alpha^2} y_2.
$$
The two solutions are asymptotically equivalent and tend to the unregularised solution as $\alpha \rightarrow 0$.
In particular, the first coefficient is exactly right with the rr-QR factorisation but not with the Tikhonov regularisation.

\subsubsection{Integral functionals}
We now introduce a class of regularisers that are made possible by the fact that
we decompose the PES into relatively low-dimensional components, the
body-orders. A special case, discussed in the next section will be a critical
ingredient in our fitting procedure.

Consider a PES given by a body-order expansion \eqref{eq:bo-expansion}, and
let us assume that we write $E_n$ as a distance-based or distance-angle potential,
i.e., $E_n(\{\bR_i\}_i) = V_n({\bf u}_n(\{\bR_i\}))$. Then we consider a
regularisation functional of the form,
\begin{equation} \label{eq:reg_intfcnl}
   \|\Gamma_n {\bf c} \|^2_2
   = \int w({\bf u}) \Big| L\big[V_n({\bf u})\big]\Big|^2 \, d{\bf u},
\end{equation}
where $L$ is a linear differential operator, $w$ an integration weight
and integration is taken over the domain of definition of $V_n$, i.e.,
all admissible tuples ${\bf u}$ that can be written as ${\bf u} = {\bf
  u}_n(\{\bR_i\})$.  The right-hand side can be written in the form of
a Tikhonov functional since $V_n$ depends linearly on a subset of
coefficients ${\bf c}$.

To approximately evaluate this integral we choose integration points
$\{{\bf u}_j\}_{j = 1}^J \subset \R^d$ that are distributed according
to the measure $w({\bf u}) d{\bf u}$ and replace the integral functional
\eqref{eq:reg_intfcnl} with its discretised variant
\begin{equation}
   \label{eq:reg_sumfcnl}
   \|\Gamma_n {\bf c} \|^2_2 = \frac{1}{J} \sum_{j = 1}^J
   \Big| L\big[V_n({\bf u}_j)\big]\Big|^2.
\end{equation}
A canonical choice for $\{{\bf u}_j\}_{j = 1}^J$ are low-descrepancy
sequences; we simply use the classical Sobol sequence. This is effective
in low and moderate dimensions where Sobol sequences ``fill space'' with
few (O(1000) to O(100,000)) points~\cite{Niederreiter-Sobol}. In principle
one could also use random number sequences instead.

\begin{figure}
   \hspace{-10mm} \includegraphics[width=8.8cm]{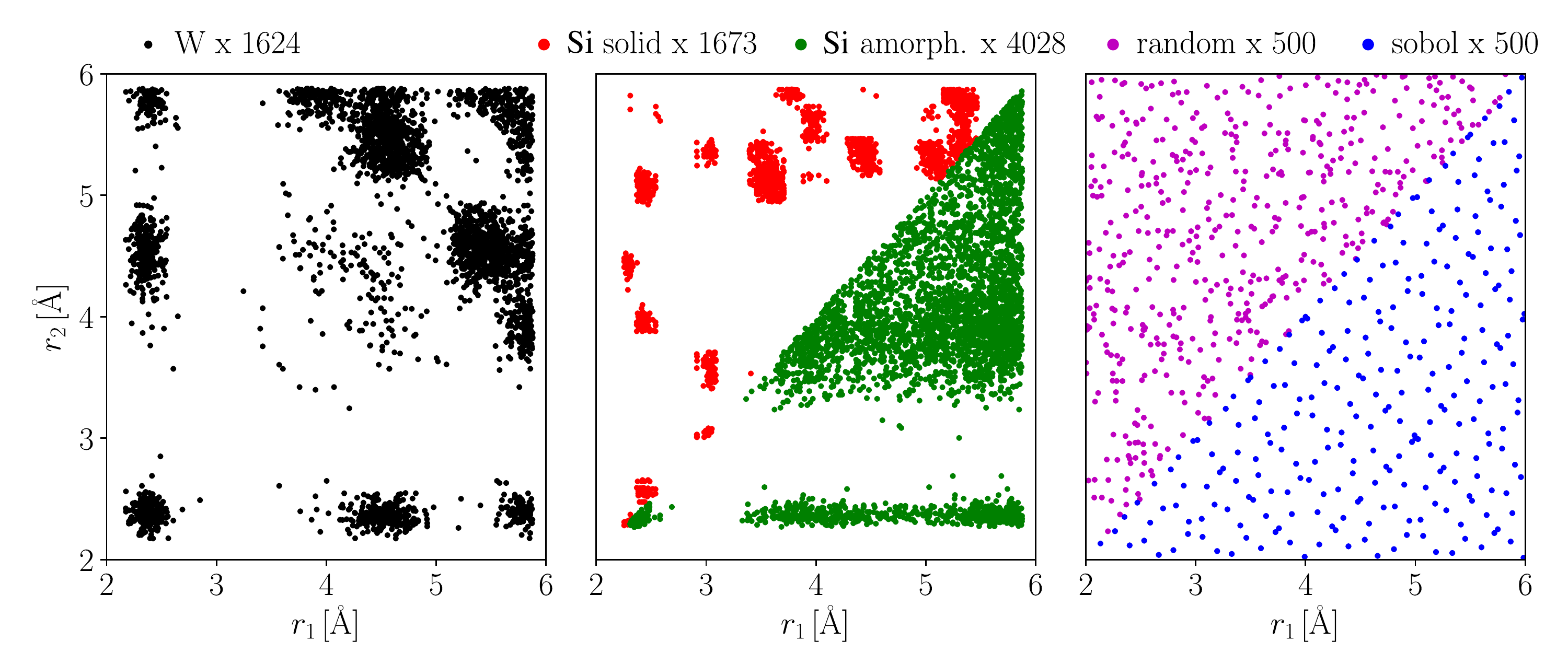}
   \caption{Distribution of 3-body clusters in the W and Si training sets versus
   artificially generated (random, Sobol) clusters.The clusters are
   represented by distance-angle variables $(r_1, r_2, w_{12})$ with
   the angle variable $w_{12} = \hat{R}_1 \cdot \hat{R}_2$
   restricted to $-0.3 \leq w_{12} \leq -0.2$. The picture is similar for
   other angle-intervals. }
   \label{fig:dbvssobol}
\end{figure}

In Figure~\ref{fig:dbvssobol} we show how solid configurations are highly
concentrated in the space of $n$-body clusters. Amorphous configurations ``fill
space'' much better (liquid even more so), and can to a certain degree be seen
as a ``natural'' regulariser, however they still concentrate in parts of
configuration space. By contrast, random or Sobol sequences provide close to
uniform distributions of datapoints at which to apply the regularisation, or
alternatively their concentration can be easily tuned by adjusting the upper and
lower bounds or through applying a distance transformation.

\subsubsection{Laplace smoother}
A wide variety of choices for the differential operator $L$ in
\eqref{eq:reg_sumfcnl} are possible. In the present work we will only consider
the Laplace operator, $L = \gamma \Delta \equiv \gamma \nabla^2$, i.e., \eqref{eq:reg_sumfcnl} becomes
\begin{equation}
   \label{eq:reg_sumfcnl_laplace}
   J_n^\Delta = \frac{\gamma_n}{J} \sum_{j = 1}^J
   \Big| \Delta \big[V_n({\bf u}_j)\big]\Big|^2,
\end{equation}
where $\gamma_n$ is an adjustable regularisation parameter.
Although it is in principle possible to implement second derivatives of $V_n$,
we have chosen instead to approximate $\Delta V_n$ with a finite-difference,
\[ \begin{split}
   \Delta_h V_n({\bf u}) =
   h^{-2} \sum_{\alpha = 1}^d \Big(  V_n({\bf u} + h {\bf e}_\alpha)
       - 2 V_n({\bf u})&  \\[-2mm]
        + V_n({\bf u} - h {\bf e}_\alpha) & \Big)
\end{split} \]
Regularising the least squares fit with the functional $J_n^\Delta$ promotes
that the {\em curvature} $\Delta V_n$ is moderate, which is a gentle
requirement of smoothness. By adjusting the parameter $\gamma_n$, smoothness can
be traded against accuracy of fit.

\subsubsection{Two-sided cutoffs and repulsive core}
\label{sec:2s+rep}
For distances well below the nearest-neighbour distance
regularisation is particularly crucial as demonstrated
in~\cite{Nandi2019-bl} and in Figure~\ref{fig:rough_vs_reg_2b2}. While
we could apply the Laplace regulariser in this region to control
oscillations of the polynomials and thus prevent ``holes'' in the PES,
this would inhibit the ability of the polynomials to produce an
accurate fit in regions of interest. Instead, we chose to (1) apply an
inner cutoff to all $V_n$, $n \geq 3$; and (2) replace the global
two-body polynomial $V_2$ with a spline that guarantees repulsion at
short interatomic distances. In detail we apply these ideas as follows:

\paragraph{Two-sided cutoff: }
Typically, the cutoff function $f_{\rm cut}$ appearing in
\eqref{eq:fcut_BL} and \eqref{eq:fcut_BA} are positive on an interval
$[0, r_{\rm cut})$ and vanish on $[r_{\rm cut}, \infty)$. For example,
    we often use the spline defined in \eqref{eq:poly_cutoff}.

\begin{figure}
   \includegraphics[width=0.75\linewidth]{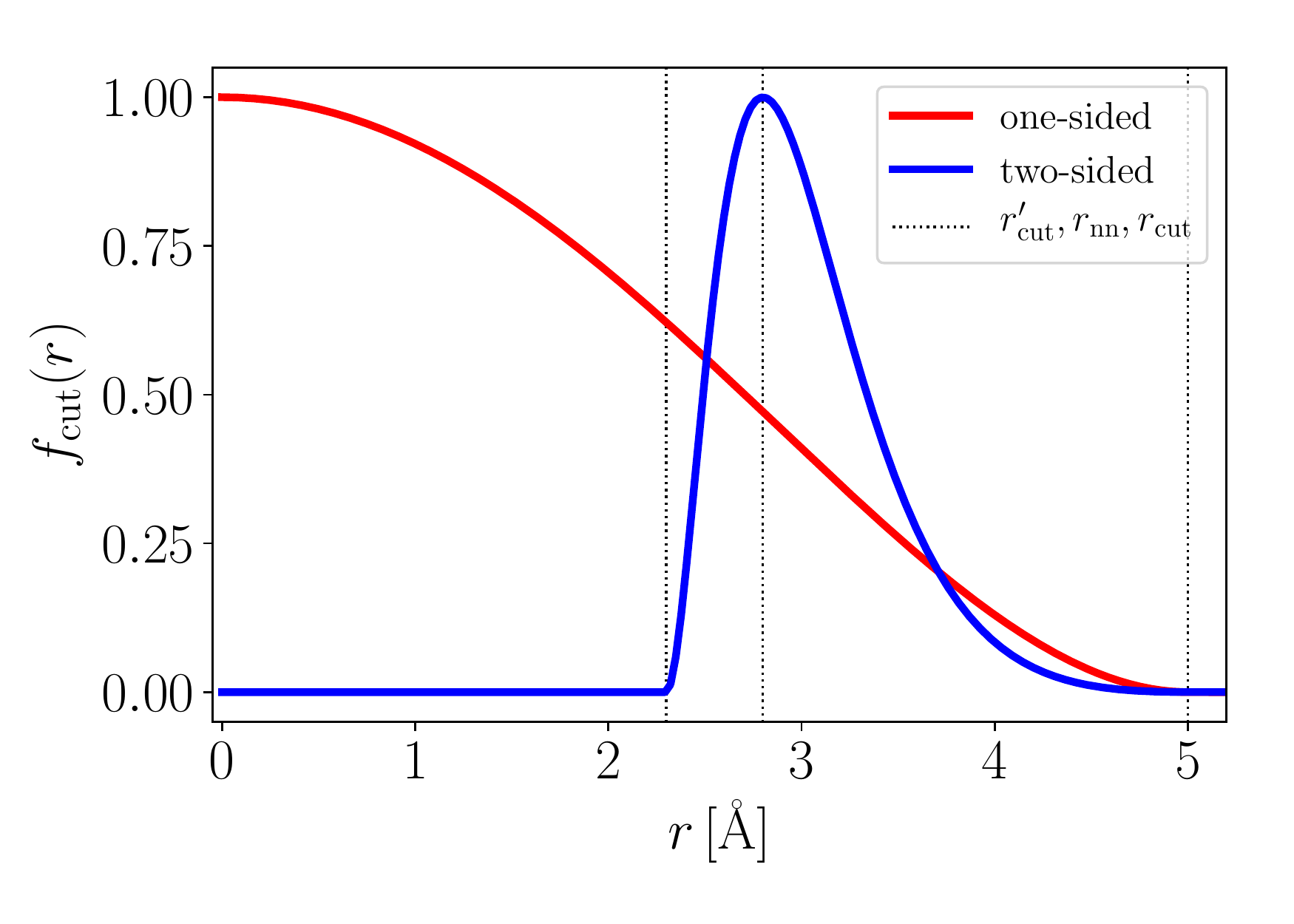}
   \caption{A one-sided cutoff function \eqref{eq:poly_cutoff} for the two-body function $V_2$ and a two-sided cutoff function \eqref{eq:2scutoff} for $n$-body functions $V_n$, $n \geq 3$. The illustrative parameters are $r_{\rm cut}' = 2.3 \text{\AA}, r_{\rm nn} = 2.8\text{\AA}, r_{\rm cut} = 5.0 \text{\AA}$.}
   \label{fig:cutoffs}
\end{figure}

In order to prevent oscillation and blow-up of the $n$-body functions
$V_n$, $n \geq 3$ we require that $f_{\rm cut} = 0$ on both $[0,
  r_{\rm cut}']$ and $[r_{\rm cut}, \infty)$. A specific choice that
  we used in our tests is
\begin{equation}
      \label{eq:2scutoff}
    \begin{split}
       f_{\rm cut}(r) &=
       \begin{cases}
             C \big( \xi^2 - 1 \big)^2, & r_{\rm cut}' < r < r_{\rm cut}, \\
             0, & \text{otherwise},
        \end{cases} \\
       \xi &= \exp\Big( \lambda \big(r/ r_{\rm nn} - 1\big) \Big) - 1,
    \end{split}
\end{equation}
where $r_{\rm nn}$ is an estimate for the ground state nearest
neighbour distance in the material under consideration, $\lambda$ is
chosen such that the resulting $f_{\rm cut}$ has its unique local
maximum at $r_{\rm nn}$ and $C$ such that $f_{\rm cut}(r_{\rm nn}) =
1$. See Figure~\ref{fig:cutoffs} to visualise this construction.
 We emphasize, however, that there are many reasonable alternatives
 to implement this.

\begin{figure}
   \includegraphics[width=\linewidth]{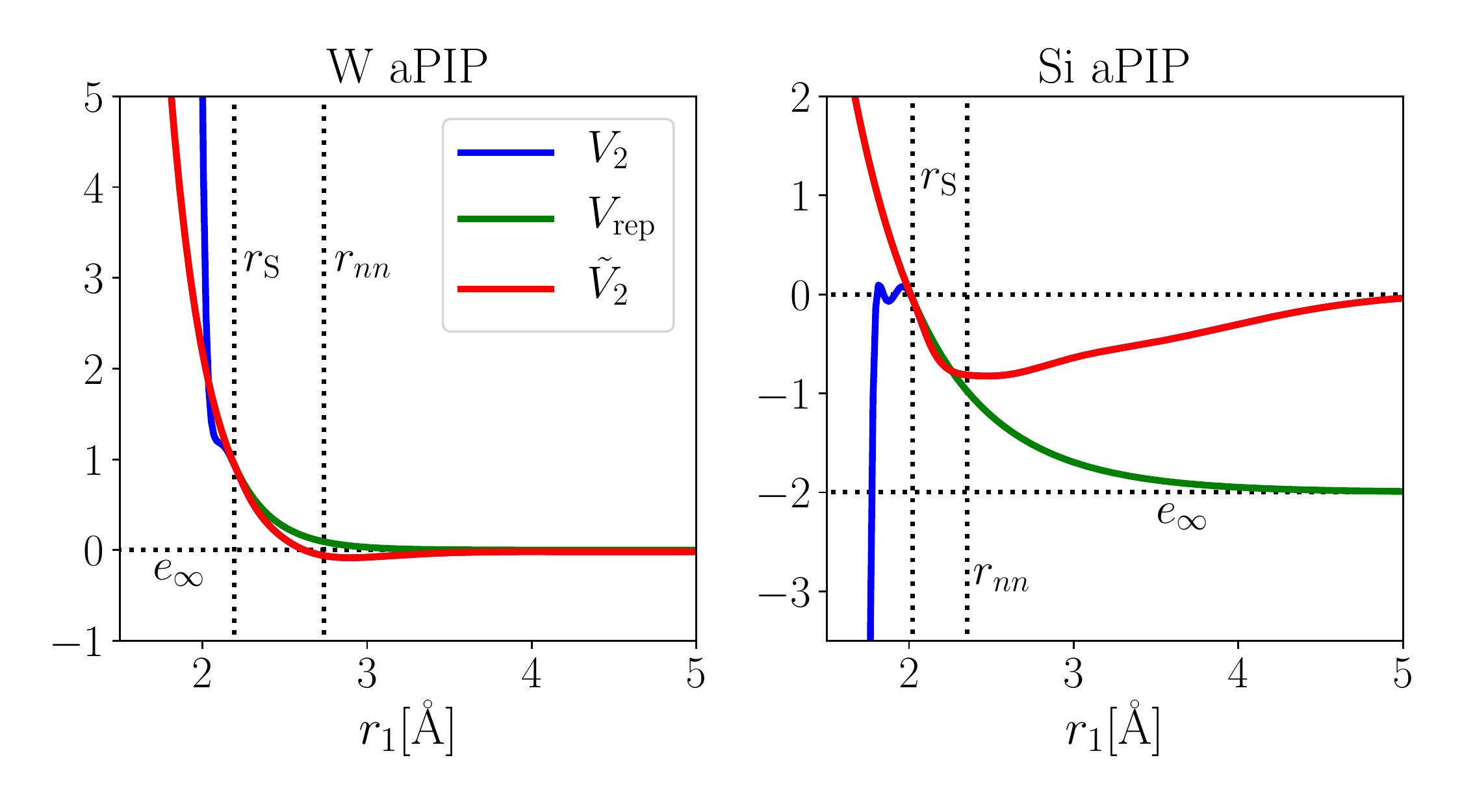}
   \caption{Spline two-body potentials with repulsive core; cf. \S~\ref{sec:2s+rep}(b).}
   \label{fig:V2rep}
\end{figure}

\paragraph{Repulsive Two-Body: }
We initially perform the regularised least-squares fit with a global
polynomial representation of $V_2$ as described in
\S~\ref{sec:pots}. We then choose a spline point $r_{\rm S} < r_{\rm
  nn}$, sufficiently small so that modifying $V_2(r)$ for $r < r_{\rm
  S}$, ideally chosen small enough so that the RMSEs are not
significantly affected. This point must furthermore be chosen so that
$V_2'(r_{\rm S}) < 0$. We then define a new two-body potential
\begin{align*}
   \tilde{V}_2(r) &:= \begin{cases} V_2(r), & r \geq r_{\rm S},
     \\ V_{\rm rep}(r), & r < r_{\rm S},
   \end{cases} \\
   V_{\rm rep}(r) &= e_\infty + \beta r^{-1} e^{- \alpha r},
\end{align*}
where $e_\infty < V_2(r_{\rm S})$ is a tuning parameter that can be
used to adjust the steepness of the potential, while $\alpha, \beta$
are chosen such that $\tilde{V}_2$ is continuous and continuously
differentiable at $r_{\rm S}$.
The form of the repulsive potential $V_{\rm rep}$ is arbitrary, and in
applications where it is important to accurately describe  interactions between
 atoms at very
close distances it should be chosen as or similar to the universal ZBL
function~\cite{Ziegler1985-sd}.
The repulsive core constructions for the regularised W and Si
fits, described in detail in \S~\ref{sec:regularisation:fitaccuracy},
are visualised in Figure~\ref{fig:V2rep}.

In practice, the inner cutoff and splining mechanisms interact mildly with the
regularised least squares regression, and we did not find it
particularly difficult to find suitable parameter choices.

\subsubsection{Sequential Fits}
A source of ill-conditioning in the least squares system
\eqref{eq:LSQ_system} is due to the fact that any $m$-body function
$V_m$ can {\em nearly} be represented as an $n$-body function $V_n$
with $n > m$. Thus, a final mechanism that we employ is to fit
different $n$-body terms independently from one another. For example,
we may first fit a two-body $V_2$, followed by the modification
described in \S~\ref{sec:2s+rep}(2). Then, in a separate step we fit
$V_3, V_4$, and possibly $V_5$ after subtracting the values for $V_2$
from the observations vector. At present we perform this procedure in
a purely {\it ad hoc} fashion.

\subsection{Accuracy of Regularised aPIPs}
\label{sec:regularisation:fitaccuracy}
In this section, we investigate the effect of our regularisation procedures on
the fit accuracy for the W and Si training sets described in
\S~\ref{sec:convergence_rmse}, as well as a small selection of material properties.
While the Si fits will be performed with a 5-body potential, for the W potential
we restrict  the body-order to four since this already achieves satisfactory
accuracy on the W training set.

The majority of hyperparameters for the Si and W fits are identical and can be
summarized as follows: All potentials are distance-angle potentials with the
same distance transforms $u_{ij}$ as in the RMSE convergence tests. For the
unregularised aPIPs the cutoff function is \eqref{eq:poly_cutoff} while for the
regularised aPIPs we use the one-sided cutoff \eqref{eq:poly_cutoff} only for
the two-body potential but a two-sided cutoff \eqref{eq:2scutoff} for all $V_n,
n \geq 2$. The least squares functionals are weighted differently from the RMSE
convergence tests where we were targeting {\em total RMSEs}. For the
regularisation and extrapolation tests we chose the weights to aim for accuracy
on subsets comparable to the previously published SOAP-GAP models. The regularised fits employ the
complete range of tools introduced in \S~\ref{sec:regularisation}. The specific
details of the aPIP potential parameters, fitting parameters and regularisation
parameters, and sequential fitting procedure, are given in the supplement.

The resulting unregularised and regularised potentials will, respectively, be
denoted by {\tt aPIP(unreg)} and {\tt aPIP(reg)}.

\subsubsection{Results: Tungsten}
\label{sec:wregfits}
We compare the aPIP RMSE for energies, forces and virials per
configuration type in the tungsten training set against the original SOAP-GAP model published with the data set~\cite{Szlachta2014-tb}.
The results can be found in Table \ref{tbl:W_errors_reg}. The purpose of this
test is to confirm that the regularisation only slightly reduces the RMSE
accuracy per atom compared to the unregularised aPIP(unreg).

\begin{figure}[ht!]
    \includegraphics[width=0.9\columnwidth]{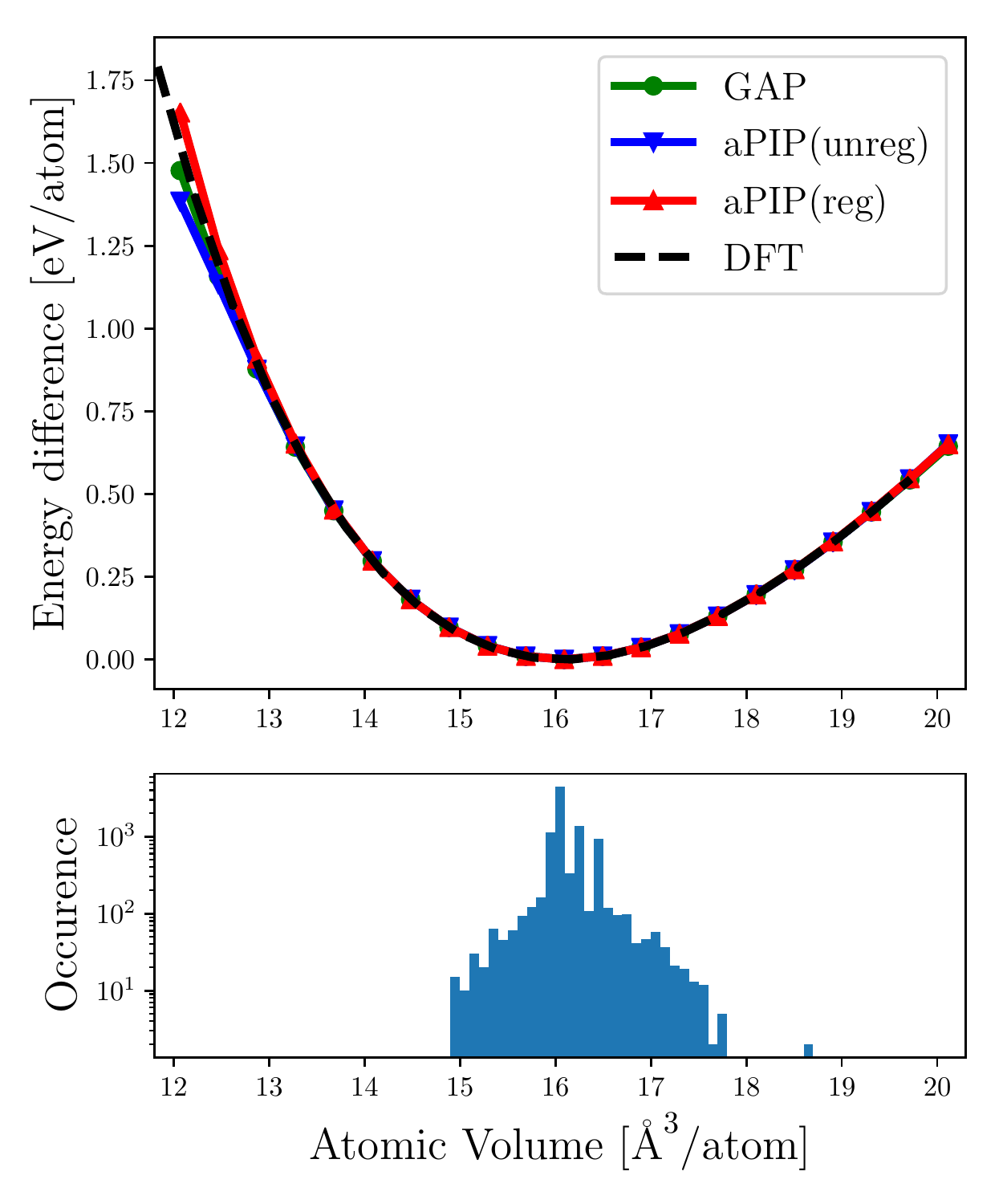}
    \caption{
    Tungsten bcc energy versus volume comparing SOAP-GAP~\cite{Szlachta2014-tb} against regularised and unregularised aPIP. The lower panel shows the histogram of volumes in the data set.}
    \label{fig:w_ev_plot}
\end{figure}

Apart from monitoring the RMSE  we also benchmark the fitted potentials by
comparing their predictions for various material  properties:
energy-volume curves of the aPIP models as well as the SOAP-GAP model are compared
against DFT in Figure~\ref{fig:w_ev_plot}. These results are shown to be in
excellent agreement for all the fitted potentials. A second test is to calculate
the elastic constants $B$, $C_{11}$, $C_{12}$, $C_{44}$ for a bcc tungsten
structure. The aPIP(unreg), aPIP(reg) and SOAP-GAP all achieve to predict the elastic constants
within 1$\%$.

\begin{table*}
   \begin{tabular}{r|ccc|ccc|ccc} \toprule[1pt]
& \multicolumn{3}{c|}{Energy [meV]} & \multicolumn{3}{c|}{Forces [meV/\AA]} & \multicolumn{3}{c}{Virials [meV]} \\config type & GAP & aPIP(unreg) & aPIP(reg) & GAP & aPIP(unreg) & aPIP(reg) & GAP & aPIP(unreg) & aPIP(reg) \\
\midrule[1pt] unit cells & 0.07 & 0.24 & 0.27 & 0.00 & 0.00 & 0.00 & 3.47 & 3.80 & 4.17\\
bulk MD & 0.60 & 0.47 & 0.45 & 27.8 & 21.4 & 26.5 &   &   &  \\
vacancy & 0.50 & 0.26 & 0.67 & 29.4 & 25.3 & 29.5 &   &   &  \\
dislocation & 1.86 & 0.98 & 1.02 & 38.3 & 33.0 & 35.7 &   &   &  \\
surface & 0.45 & 0.43 & 0.88 & 50.9 & 42.2 & 70.8 &   &   &  \\
$\gamma$-surface & 1.66 & 2.92 & 4.09 & 69.0 & 99.4 & 118.0 &   &   &  \\
$\gamma$-s. vacancy & 1.26 & 1.70 & 2.71 & 79.3 & 98.7 & 109.3 &   &   &  \\
\toprule[1pt] \end{tabular}
   \caption{RMSE Accuracy on W training set, comparing SOAP-GAP~\cite{Szlachta2014-tb} against regularised and unregularised aPIP.
   }
   \label{tbl:W_errors_reg}
\end{table*}

\subsubsection{Results: Silicon}
\label{sec:siregfits}
The RMSE accuracy per configuration type of unregularised aPIP(unreg) and
regularised aPIP(reg) are compared against the SOAP-GAP fit\cite{Bartok:2018ih} in Table
\ref{tbl:Si_errors_reg}. Again the regularised aPIP(reg) is shown to decrease in
RMSE accuracy compared to the unregularised aPIP(unreg), and by larger factors than in the case of tungsten.

\begin{table*}
   \begin{tabular}{r|ccc|ccc|ccc} \toprule[1pt]
& \multicolumn{3}{c|}{Energy [meV]} & \multicolumn{3}{c|}{Forces [meV/\AA]} & \multicolumn{3}{c}{Virials [meV]} \\config type & GAP & aPIP(unreg) & aPIP(reg) & GAP & aPIP(unreg) & aPIP(reg) & GAP & aPIP(unreg) & aPIP(reg) \\
\midrule[1pt] dia & 0.65 & 0.33 & 0.55 & 15.4 & 13.9 & 21.8 & 14.59 & 6.09 & 9.96\\
amorph & 0.56 & 2.08 & 4.49 & 102.7 & 138.6 & 172.0 & 60.9 & 15.85 & 31.32\\
bt & 0.72 & 0.27 & 0.45 & 17.2 & 18.2 & 31.2 & 26.47 & 12.94 & 22.15\\
vacancy & 0.54 & 0.36 & 0.71 & 48.3 & 46.2 & 62.7 & 9.82 & 7.52 & 11.89\\
sp2 & 0.48 & 0.70 & 1.82 & 48.6 & 45.0 & 79.2 &   &   &  \\
surface110 & 0.20 & 0.48 & 2.80 & 161.2 & 159.0 & 237.0 &   &   &  \\
surface111 & 0.22 & 0.31 & 0.99 & 156.9 & 155.8 & 233.7 &   &   &  \\
surface001 & 0.19 & 0.50 & 1.39 & 140.4 & 136.9 & 195.7 &   &   &  \\
\toprule[1pt]\end{tabular} \\
      \caption{RMSE Accuracy on Si training set, comparing SOAP-GAP~\cite{Bartok:2018ih} against regularised and unregularised aPIP.
      }
      \label{tbl:Si_errors_reg}
\end{table*}

\begin{figure}
    \includegraphics[width=0.9\columnwidth]{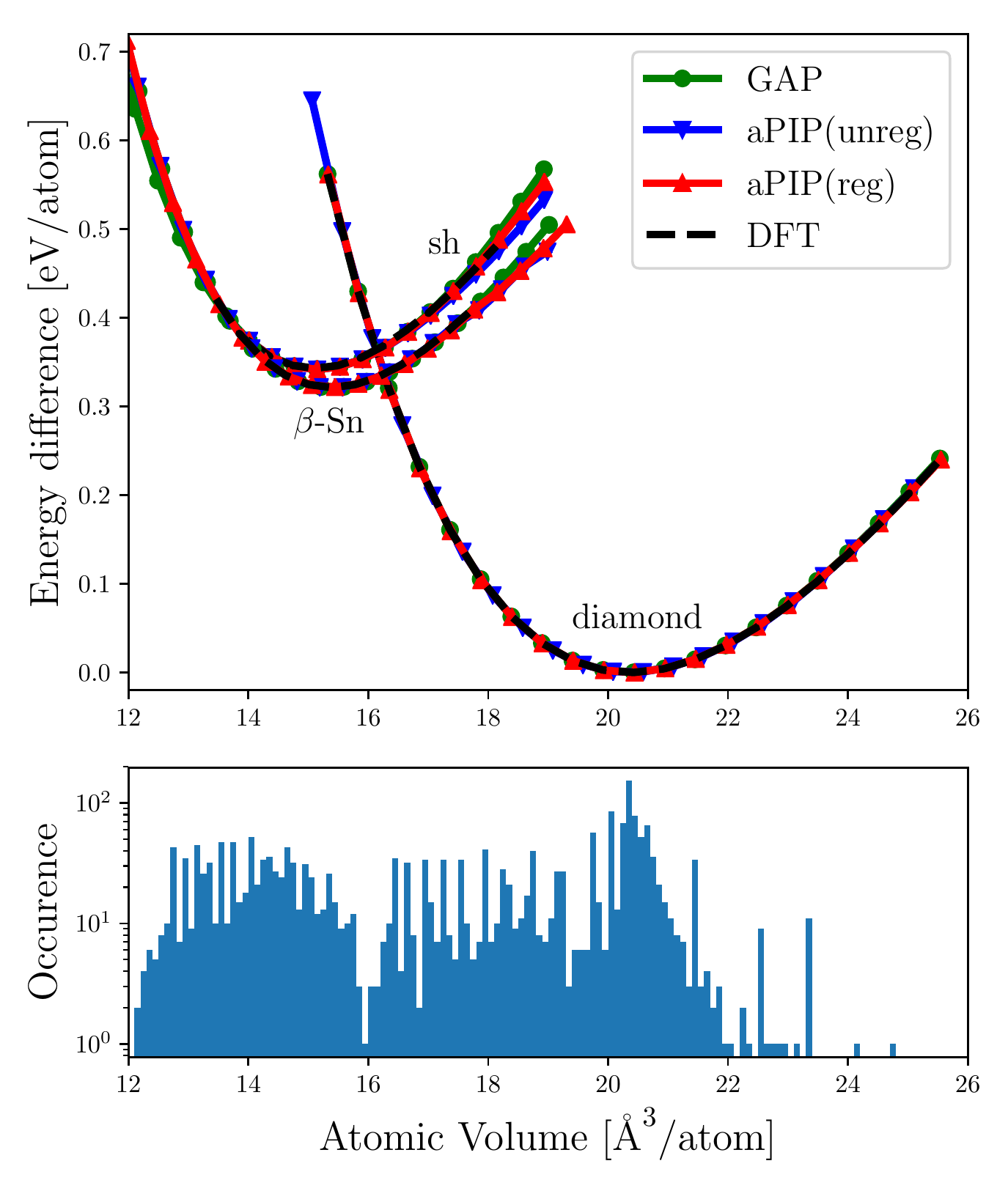}
    \caption{Energy volume curve for diamond and $\beta$-Sn silicon comparing the regularised and unregularised  aPIPs, and SOAP-GAP potentials against DFT. The lower panel shows the histogram of the volumes in the data set. }
    \label{fig:si_ev_plot}
\end{figure}

The fitted potentials were again compared for a range of different material
properties. The energy versus volume curve for silicon is shown in Figure
\ref{fig:si_ev_plot} comparing the fitted potentials to the DFT reference, with
 excellent agreement for each.
The elastic constants were calculated as well as the
surface and vacancy formation energies and are presented in Table
\ref{tbl:Si_elas_const}. The unregularised aPIP(unreg) is shown to
have larger errors on the elastic constants compared to the aPIP(reg)
and most notably failed the vacancy energy test. That is, during the
relaxation of the vacancy using the aPIP(unreg) potential the
optimiser failed to find a local minimiser. This is a manifestation of
``holes'' in the fit which will be discussed in \S~\ref{sec:holes}.  In comparison to
SOAP-GAP the aPIP(reg) performs well in calculating the elastic constants
and vacancy formation energy. The aPIP(reg) performs worse in the
surface formation energies, most specifically the (111) direction,
which we believe can only be rectified by using even higher body
order terms. Implementing this efficiently is a focus of future work.

\begin{table*}
        \begin{tabular}{r|cccc|ccc|c} \toprule[1pt]
\multirow{2}{*}{Model} & \multicolumn{4}{c|}{Elastic Constants [GPa]} & \multicolumn{3}{c|}{Surface Energy [J / m$^{2}$]} & Point defect [eV]\\
  & B & $C_{11}$ & $C_{12}$ & $C_{44}$ & (100) & (110) & (111) & vacancy \\  \toprule[1pt]
  DFT & 87.45 & 152.21 & 55.07 & 74.95 & 2.17 & 1.52 & 1.57 & 3.67 \\[-3mm]
  \multicolumn{8}{c}{} \\
  \multicolumn{9}{c}{Relative Error ($\%$)} \\ \toprule[1pt]
GAP & $<$1 & -3 & 3 & -7 & -1 & -1 & -3 & -2 \\
aPIP(unreg) & 8 & 5 & 12 & -1 & -6 & -2 & -5 & - \\
aPIP(reg) & $<$1 & 3 & -5 & -4 & -3 & -3 & -10 & -4 \\
\toprule[1pt]
\end{tabular}
        \caption{Relative error on a range of different properties for Si, comparing the SOAP-GAP, aPIP and aPIP(reg) potentials. The unregularised aPIP failed the vacancy test.}
   \label{tbl:Si_elas_const}
\end{table*}

\subsection{Holes}
\label{sec:holes}

As a first test of the ``transferability'' of our regularised aPIP
fits we determine whether the resulting potentials have any ``holes'' in
the sense of~\cite{Nandi2019-bl}: regions of unphysically low potential energy values. The importance of avoiding such behaviour is hard to overstate: in most molecular modelling applications, samples will be drawn using the {\em model} distribution, and due to the exponential amplification of the Boltzmann distribution at low and moderate temperature, holes can lead to catastrophic failure of models.

\begin{figure}
   \centering
   \includegraphics[width=\linewidth]{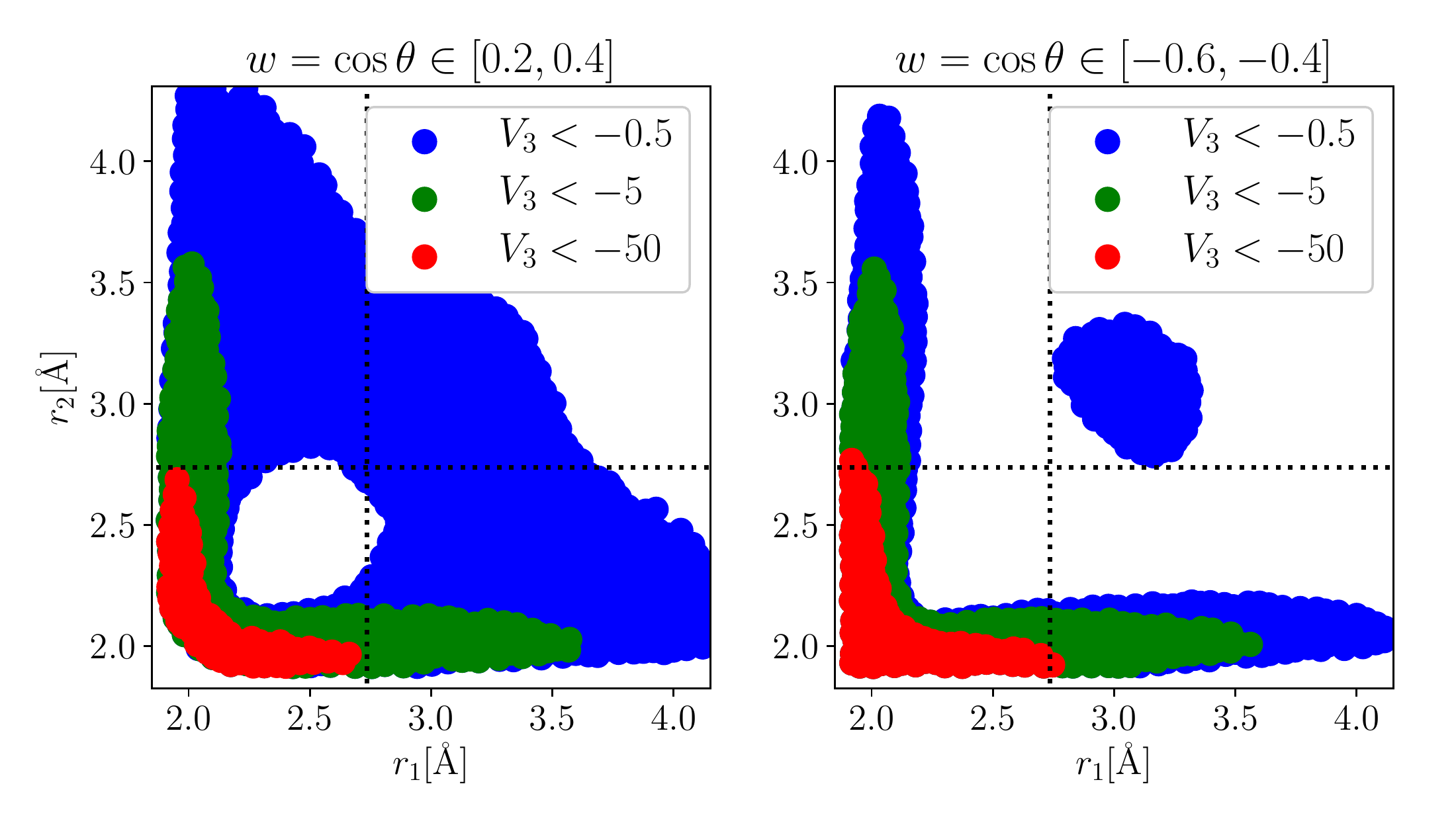}
   \caption{\label{fig:holes} Holes in the 3-body potential $V_3$ for the unregularised W aPIP fit. Two slices through the potential energy landscape are plotted.}
\end{figure}

Having decomposed the total PES into low-dimensional components gives
the option of searching for low-energy configurations in these
individual components $V_n$. Specifically, we choose a minimal inner
distance $r_0 < r_{\rm cut}'$, i.e., below the inner cutoff. Then, for
each $n$-body term $V_n$ we compute an approximate minimum of
$V_n(\bR_1, \dots, \bR_n)$ over all clusters $(\bR_1, \dots, \bR_n)$ with $r_0
\leq r_j \leq r_{\rm cut}$, using Sobol sequences with a few million
points. The results are summarized in Table~\ref{tbl:W_holes}. It is
interesting to note that the ``holes'' in the unregularised Si fit are less
severe. We speculate that this is due to the fact that the Si training
set is much richer; cf. Figure~\ref{fig:holes}.

\begin{table}
   \begin{tabular}{r|ccc}
      \toprule[1pt]
      &                & $\inf V_n$ \\
      & $~~n~~$ & ~aPIP(unreg)~ & ~aPIP(reg)~ \\
      \midrule[1pt]
      \quad W~ & 2     & -273 eV & -0.08 eV \\
        & 3     & -192359 eV & -0.06 eV \\
        & 4     & -4877 eV & -0.18 eV \\
     \midrule[1pt]
     Si~ & 2     & -0.66 eV & -0.08 eV \\
      & 3     & -2934 eV & -0.6 eV \\
      & 4     & -447 eV & -0.11 eV \\
      \bottomrule[1pt]
   \end{tabular}
   \caption{``Holes'' in the unregularised aPIP(unreg) potentials.}
   \label{tbl:W_holes}
\end{table}

In the unregularised fit we visualise the location of holes by
plotting slices through the $V_3$ energy landscape in
Figure~\ref{fig:holes}. This shows in particular that holes appear
both in regions of large angles and moderate angles near the
ground-state.

\section{Generalisation}
\label{sec:extrapolation}
This section presents a series of tests to benchmark the generalisation
performance of the aPIP and GAP models
against a DFT reference. These tests were designed to probe configurational
space far from the training data region and monitor each models' performance. We
use the term ``generalisation'' (also often called ``extrapolation'' or ``transferability'') in a loose
sense and simply take it to mean ``evaluation far from the training set'' which
may technically also include {\em interpolation}---the distinction is tenuous in
high dimension.

\subsection{Tungsten}

\subsubsection{Interstitial}
\begin{figure}
    \includegraphics[width=1.0\columnwidth]{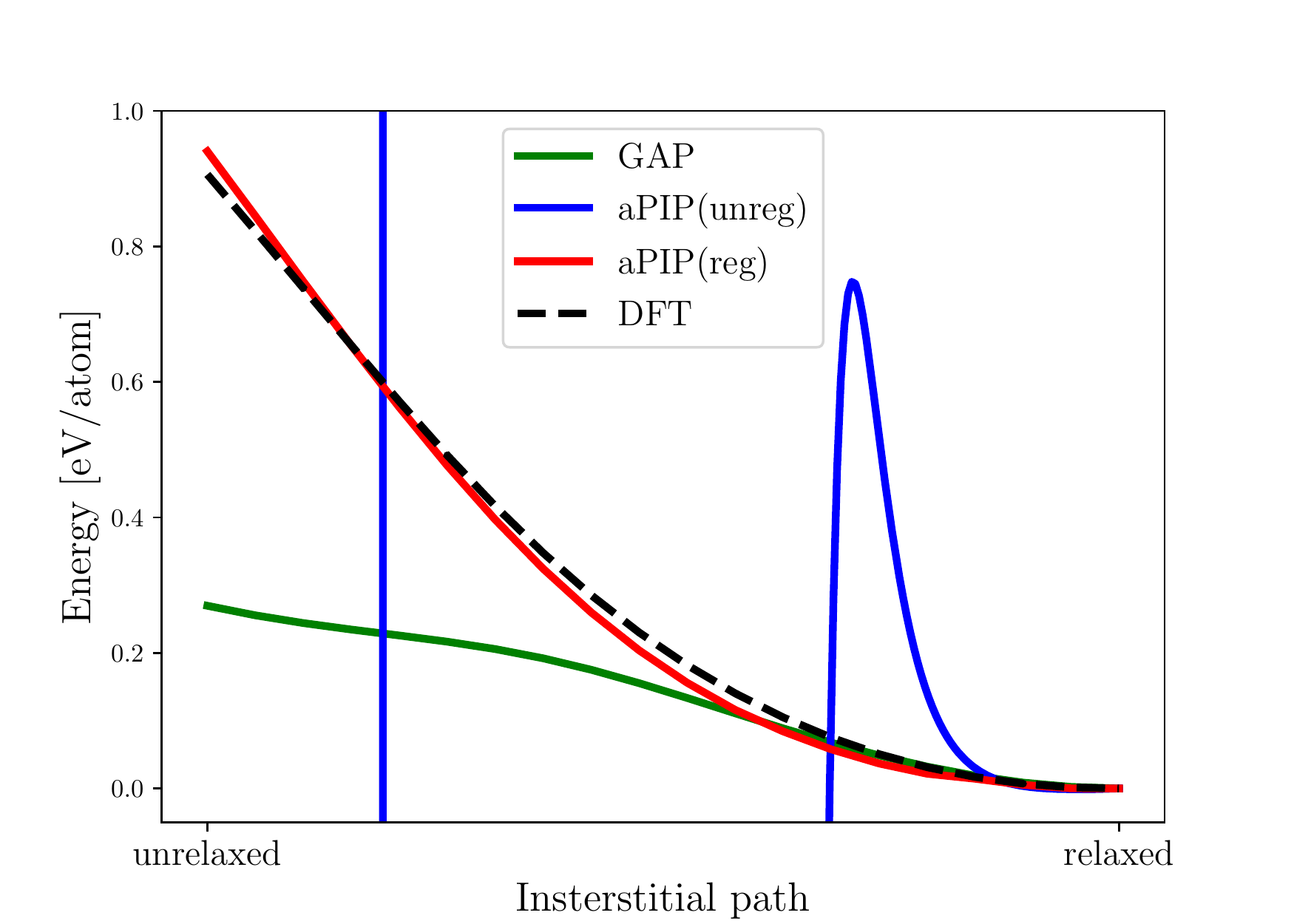}
    \caption{
    Linear path between unrelaxed and relaxed tungsten interstitial configuration. The unregularised aPIP fit explodes whereas the regularised aPIP closely matches the DFT reference.
    \label{fig:w_interstitial}
    }
\end{figure}

The tungsten aPIP(unreg) and aPIP(reg) models introduced in \ref{sec:wregfits} were compared against the
original SOAP-GAP model~\cite{Szlachta2014-tb} and a DFT reference
for the self-interstitial defect. The DFT parameters were chosen to be identical
to those used in the original paper. These settings were: 600 eV cutoff energy, 0.03 \AA$^{-1}$ kpoint spacing and 0.1 eV smearing width.
The training set does not include interstitial data; cf. Table \ref{tbl:W_errors_reg}. We formed the
interstitial defect by inserting an atom in a DFT relaxed 54 atom
tungsten bcc cell at the octahedral site ($\frac{1}{2}$, 0, 0) of the
primitive cell. The geometry was then relaxed using DFT. A linear path
between the unrelaxed  and relaxed configurations was
created and the DFT, SOAP-GAP and aPIP models were evaluated along this
path and are shown in Figure \ref{fig:w_interstitial}.

This interstitial test probes small interatomic distances which are
not contained in the training database and can therefore be strictly seen as an
extrapolation test. Due to the smoothness prior, the SOAP-GAP model
underestimates the energy difference along the interstitial path
compared to the DFT reference. As expected, the unregularised aPIP
model heavily oscillates along this test path since it explores configurations
it was not fitted to. By contrast, the
combination of integral regularisers, repulsive core and inner
cut-offs result in a regularised aPIP model that shows an excellent match to the DFT curve.
The level of agreement is likely fortuitous, but we expect that in general the correct
repulsive nature of the potential is obtained due to having enforced this property in the two body function
$V_{2}$.

\subsection{Silicon}

\subsubsection{Surface Decohesion}
We set up a bulk Si diamond 10x1x1 supercell and increase the lattice vector length in the long direction while
keeping the atomic positions fixed which in turn creates two surfaces~\cite{Bartok:2018ih}.
Both the initial bulk structure and final surfaces are configurations that are well represented in the training set
and fitted by the aPIP(unreg), aPIP(reg) and SOAP-GAP models to within 3 meV accuracy; cf. Table \ref{tbl:Si_errors_reg}.
The configurations along the path in between are not contained in the database. Therefore this test can be seen as generalising in that it evaluates the potential on a path between two accurately fitted configurations.

\begin{figure}
    \centerline{\includegraphics[width=0.9\columnwidth]{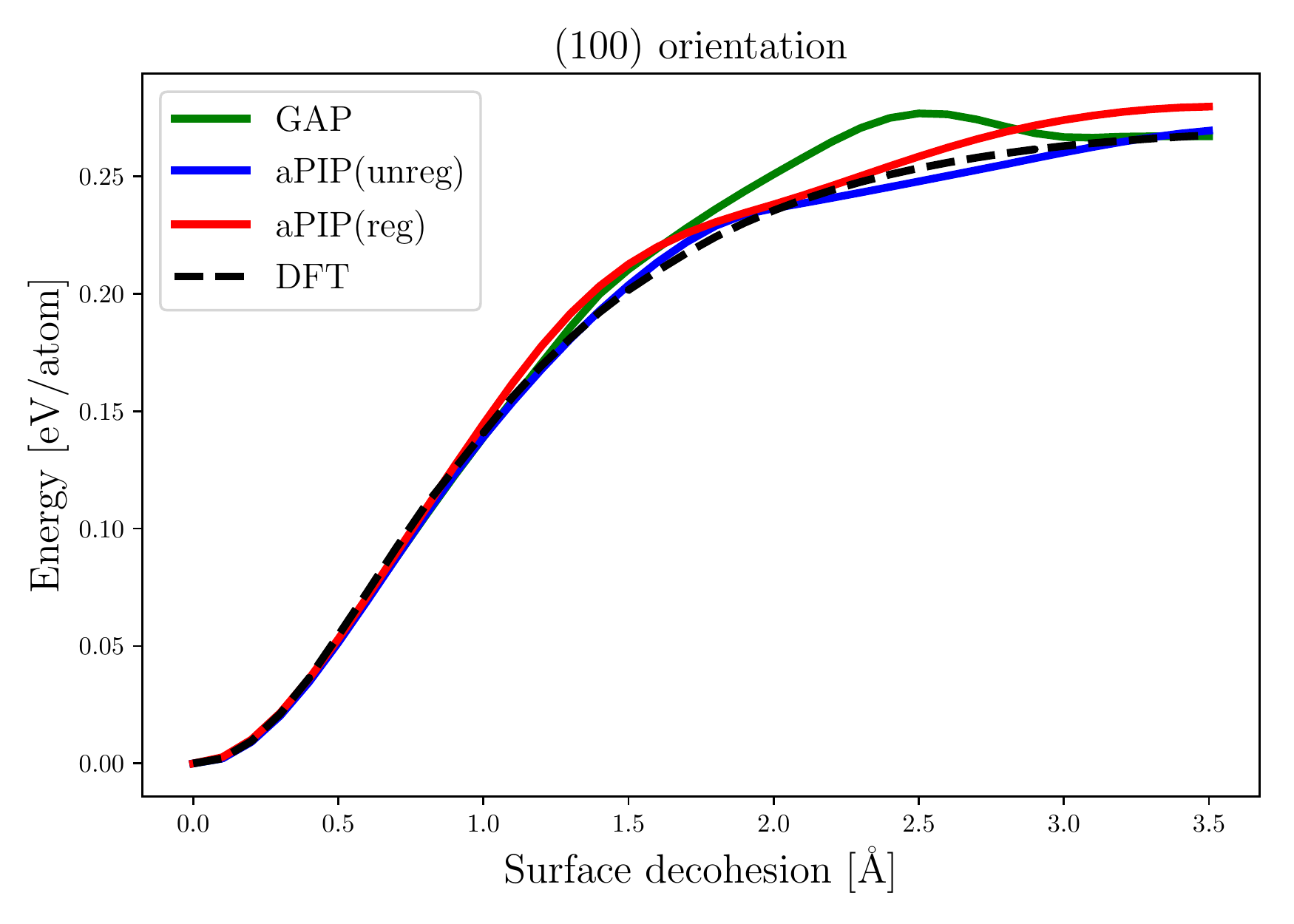}}
    \caption{Decohesion curve in silicon bulk diamond along the $(100)$ direction. The bulk diamond structure corresponds to the left end and the unrelaxed $(100)$ surface is on the right.}
    \label{fig:si_surface_decoh}
\end{figure}

Figure \ref{fig:si_surface_decoh} shows that the end points, bulk diamond and
(100) surface, are accurately fitted. However, the SOAP-GAP has a local maximum
around 2.5~\AA, unlike the DFT reference which shows a smooth and monotone
transition along this path, a characteristic which both the aPIP(unreg) and
aPIP(reg) mimic more accurately than SOAP-GAP.

\subsubsection{Layer Test}

\begin{figure}[ht]
    \centerline{\includegraphics[width=0.9\columnwidth]{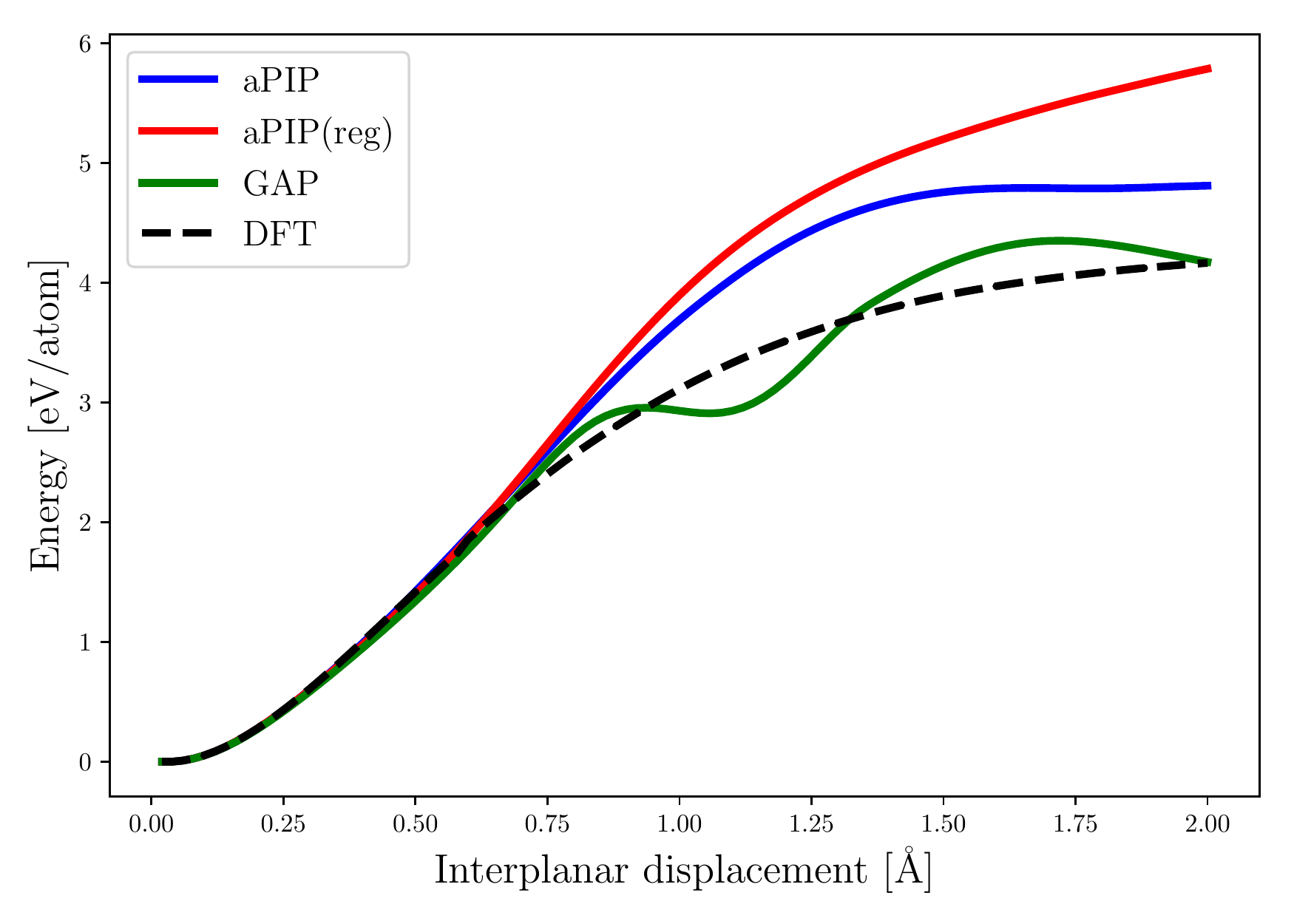}}
    \caption{The layer test along which bulk silicon is split in layers of silicene. The SOAP-GAP model predicts a high energy local minima whereas the regularised aPIP model does not.}
    \label{fig:si_layer_test}
\end{figure}

In this test we set up a bulk silicon diamond configuration and
and gradually increase the interplanar spacing between the $(111)$
layers of silicon. The configurations along this path are arguably
unphysical but should, as the DFT model confirms, correspond to
unstable high energy configurations. Past experience shows that poorly fitted potentials can have unphysically low energies for such configurations.  The aPIP and SOAP-GAP models were
evaluated on this path and are compared to the DFT reference in Figure
\ref{fig:si_layer_test}.

The SOAP-GAP model predicts a high energy local minimum along this path,
which should be entirely unstable as shown by the DFT reference. The presence
of such false high energy local minima are detrimental for applications
such as random structure search~\cite{Pickard:2011bu} but also high temperature or high stress molecular dynamics. The aPIP(unreg)
model has a much more shallow high energy local minimum, but the
aPIP(reg) model show no minimum at all (while still not being quantitatively accurate in the unphysical region).

\subsection{Titanium}
\label{sec:extrap:Ti}

Titanium is a difficult material to construct interatomic potentials for, due to its bonding chemistry being intermediate between covalent and metallic and we therefore chose it for our final test system. Here we want to explore how the different functional forms and regularisation strategies perform in the limit of {\em very little data}. The motivation for this is partly that the large size of the published data sets in the previous sections in and of itself acts like a regulariser, but also that in the future we wish to eliminate extensive sampling as a way to generate data sets. Rather, we would like to develop potential fitting frameworks that are explicitly regularised to the extent that a few judiciously chosen training configurations are sufficient to obtain good interatomic potentials.

We generated two very limited training sets, denoted by Set 1 and Set 2.
 To Set 1, we fit unregularised and regularised aPIP potentials denoted,
respectively, by ${\rm aPIP}_1 {\rm (unreg)}$ and ${\rm aPIP}_1 {\rm (reg)}$, as
well a SOAP-GAP potential denoted GAP$_1$. In addition we fit an unregularised aPIP potential,
denoted aPIP$_2$ (unreg) to Set 2. The detailed potential and fitting parameters are
described in the supplementary information.

\begin{description}
\item[Set 1] This set contains primitve cell bcc
and hcp configurations, obtained by
sampling the Boltzmann distribution with a temperature parameter set to 100~K as the lattice vectors (and in case of bcc, the relative  positions of the two atoms) are varied.  The  obtained configurations were
evaluated using CASTEP~\cite{CASTEP2005} with k-point spacing set to 0.015 \AA$^{-1}$,
750 eV cutoff energy and 0.1 eV smearing width. In addition 3x3x3 bcc and hcp supercells were added, and  Phonopy~\cite{TOGO20151} was used to generate the inequivalent finite displacements (one for each structure) of a single atom by a magnitude of 0.001~\AA. These two configurations were evaluated with a larger k-point spacing equal to 0.03 \AA$^{-1}$.

\item[Set 2] This set contains Set 1 as well as additional finite displacement configurations analogous to those in Set 1, but now with a 0.01~\AA~displacement.
\end{description}

\subsubsection{Cohesive energy}
\begin{figure}[ht]
    \includegraphics[width=0.9\columnwidth]{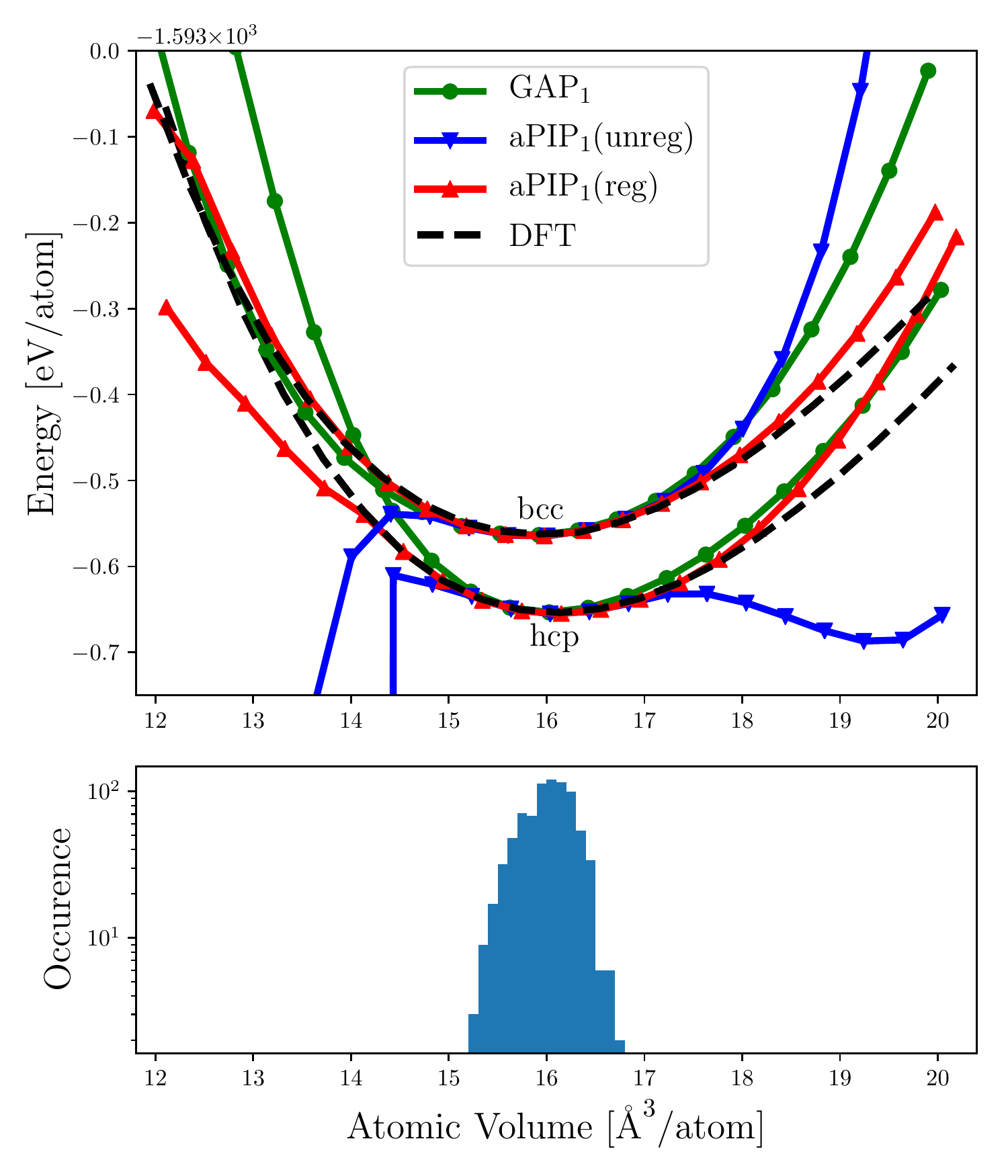}
    \caption{Energy versus volume plots for Titanium hcp and bcc phase. The volume histogram below shows the distribution of the training data.}
    \label{fig:ti_ev}
\end{figure}

Figure \ref{fig:ti_ev} shows the energy versus volume curves for the titanium bcc and hcp phases for the various potentials all fitted to Set 1. The
distribution of the atomistic volumes are on the lower panel and  show that
training data covers only volumes in the range between 15 and 17 \AA$^{3}$/atom. In this region
both aPIP models as well as the GAP model agree well with the DFT reference. For
smaller and larger volumes the fitted potentials all deviate from the
DFT reference as expected due to the lack of data. However, the unregularised
aPIP shows a steep drop for small volumes and a second minimum for large volumes whereas the regularised
aPIP and GAP mimic the DFT curve at least qualitatively, showing the benefits of regularisation.

\subsubsection{Phonon spectrum}

\begin{figure}[htp]
    \centering
    \begin{subfigure}{0.5\textwidth}
      \centering
      \includegraphics[width=0.9\columnwidth]{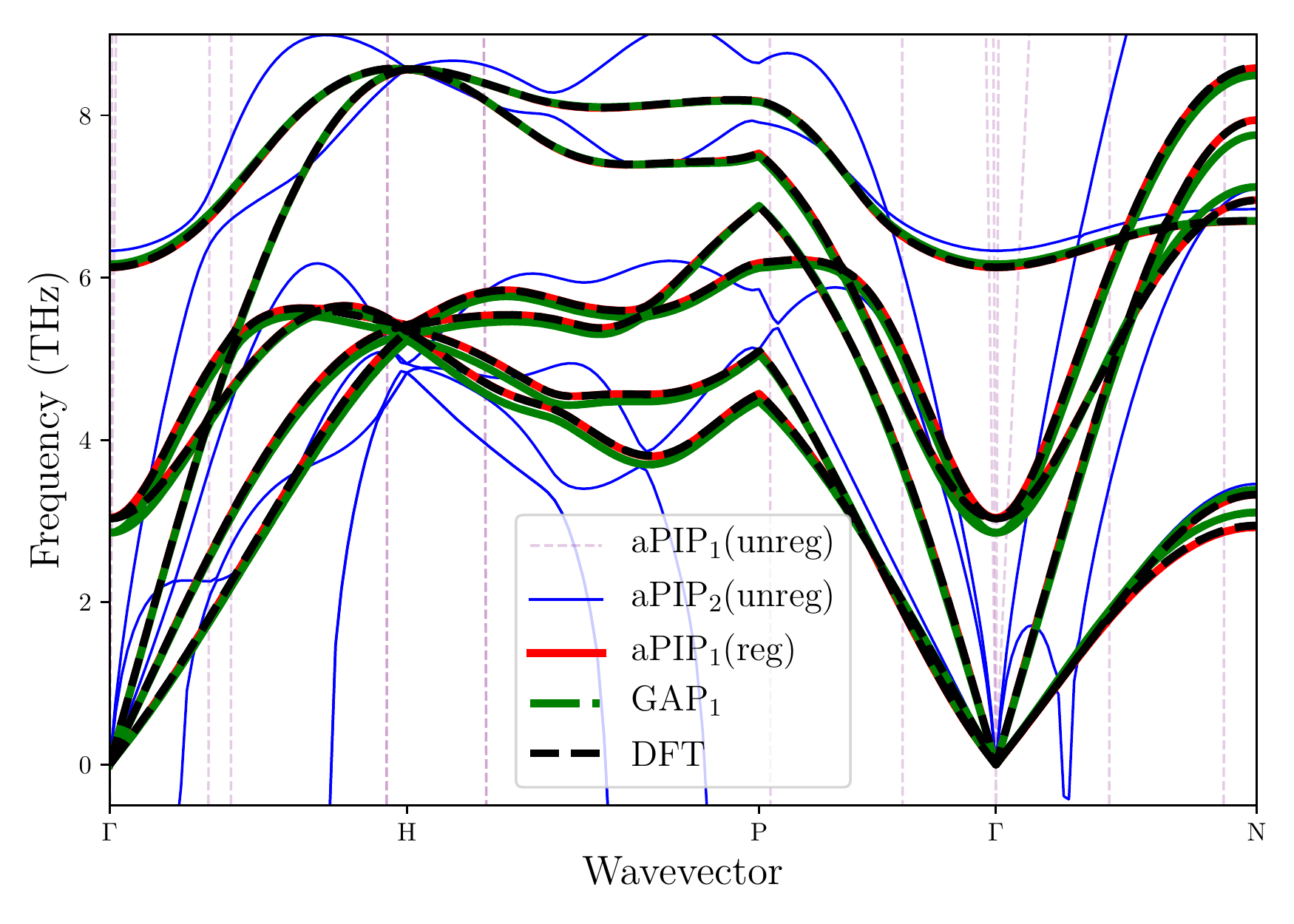}
      \caption{Titanium hcp phonon spectrum}
    \end{subfigure}%
    \\
    \begin{subfigure}{0.5\textwidth}
      \centering
      \includegraphics[width=0.9\columnwidth]{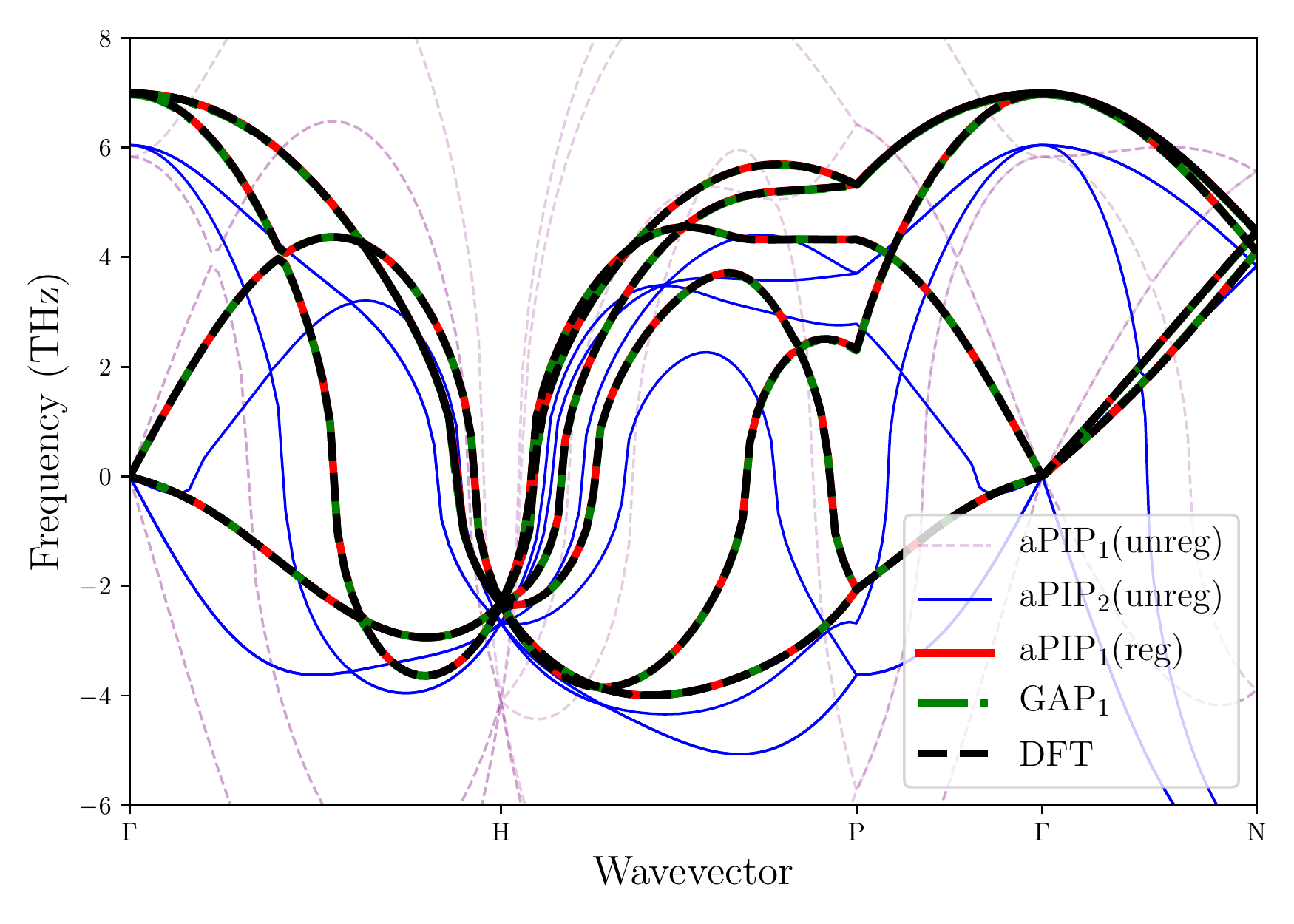}
      \caption{Titanium bcc phonon spectrum}
    \end{subfigure}
  \caption { Phonon spectra computed using the aPIP, SOAP-GAP models and DFT reference. Regularisation
  increases the accuracy of the aPIP models, similarly to the addition of FDC data.}
  \label{fig:ti_phonon}
\end{figure}

SOAP-GAP and aPIP models fitted to Set 1 and Set 2 were used to generate the phonon spectra for
bcc and hcp titanium and are compared to the DFT reference in Figure
\ref{fig:ti_phonon}.
Although aPIP$_{1}$(unreg) and aPIP$_{1}$(reg) were both fitted to Set 1, aPIP$_{1}$(unreg) failed catastrophically while aPIP$_{1}$(reg) produces a highly accurate
phonon spectrum. The negative frequencies of the bcc phonon spectrum at the $\Gamma$ points demonstrates the instability of the structure at 0~K.

Of course an alternative, but much less controllable way to improve the regularity of a potential is to simply add more data to the training set. ML potentials in the literature, even if they are not explicitly regularised, can avoid the unphysical behaviour seen in here because they are fitted to large data sets. To see this effect, consider the potentials fitted to Set 2:  aPIP$_{2}$(unreg) phonon spectra are somewhat improved, but they remain highly inaccurate quantitatively. We expect that much more training data would be required to accurately converge the phonon spectrum of an unregularised potential.  We suggest instead that regularisation and a single displacement per crystal structure are enough to accurately reproduce phonon spectra.

\subsubsection{Burgers' path}
The Burgers' path is a pathway from the bcc to hcp crystal
structure \cite{burgers1934process}. It consists of a shear
deformation applied to the bcc structure followed by a shuffle of atomic
layers resulting in a hcp structure \cite{caspersen2004importance}.

\begin{figure}
    \centerline{\includegraphics[width=1.0\columnwidth]{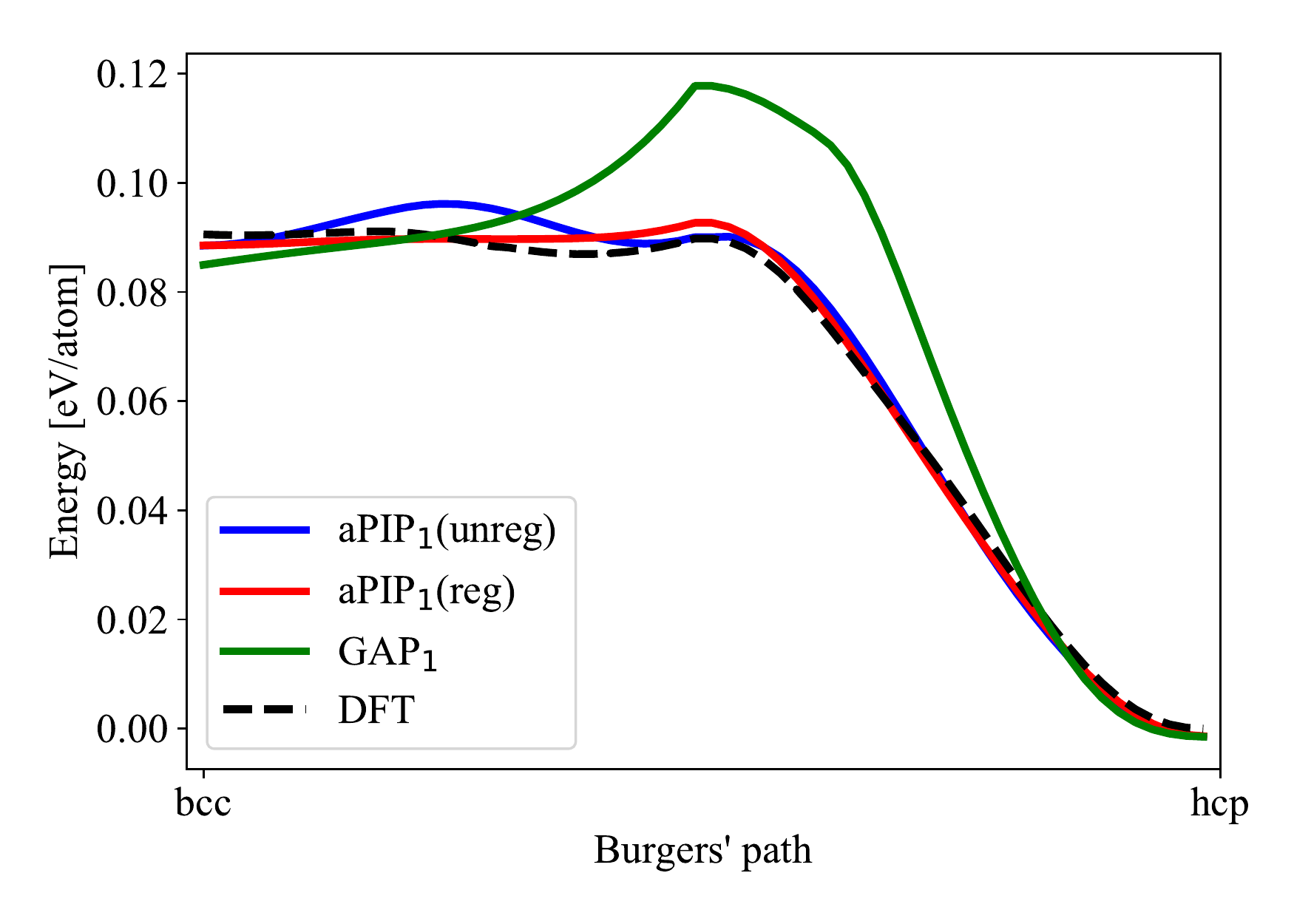}}
    \caption{Burgers' path comparing aPIP, SOAP-GAP and DFT reference. The SOAP-GAP model correctly predicts the relaxed bcc/hcp energies at both ends of the path but overestimates the energy along the remainder of the path. }
    \label{fig:ti_burgers_path}
\end{figure}

We evaluated the Set 1 fits on the Burgers' path and the results are
plotted against the DFT reference in Figure \ref{fig:ti_burgers_path}.
The energy per atom along the Burgers' path shows that the
SOAP-GAP model overestimates the barrier by 30 meV along the transition from bcc to hcp. The aPIP models
as well as the DFT reference do not show such a barrier. The training database contains
bcc/hcp configurations sampled at a low temperature (T=100~K) resulting in a database containing structures close
to the relaxed bcc and hcp configurations. As expeced, Figure \ref{fig:ti_burgers_path} shows the
aPIP and SOAP-GAP models both predict the energies for the hcp and bcc crystals accurately.
However, the aPIP model manages to predict the energy along the middle part
of the path more accurately compared to the DFT reference.

\begin{figure}
    \centerline{\includegraphics[width=0.85\columnwidth]{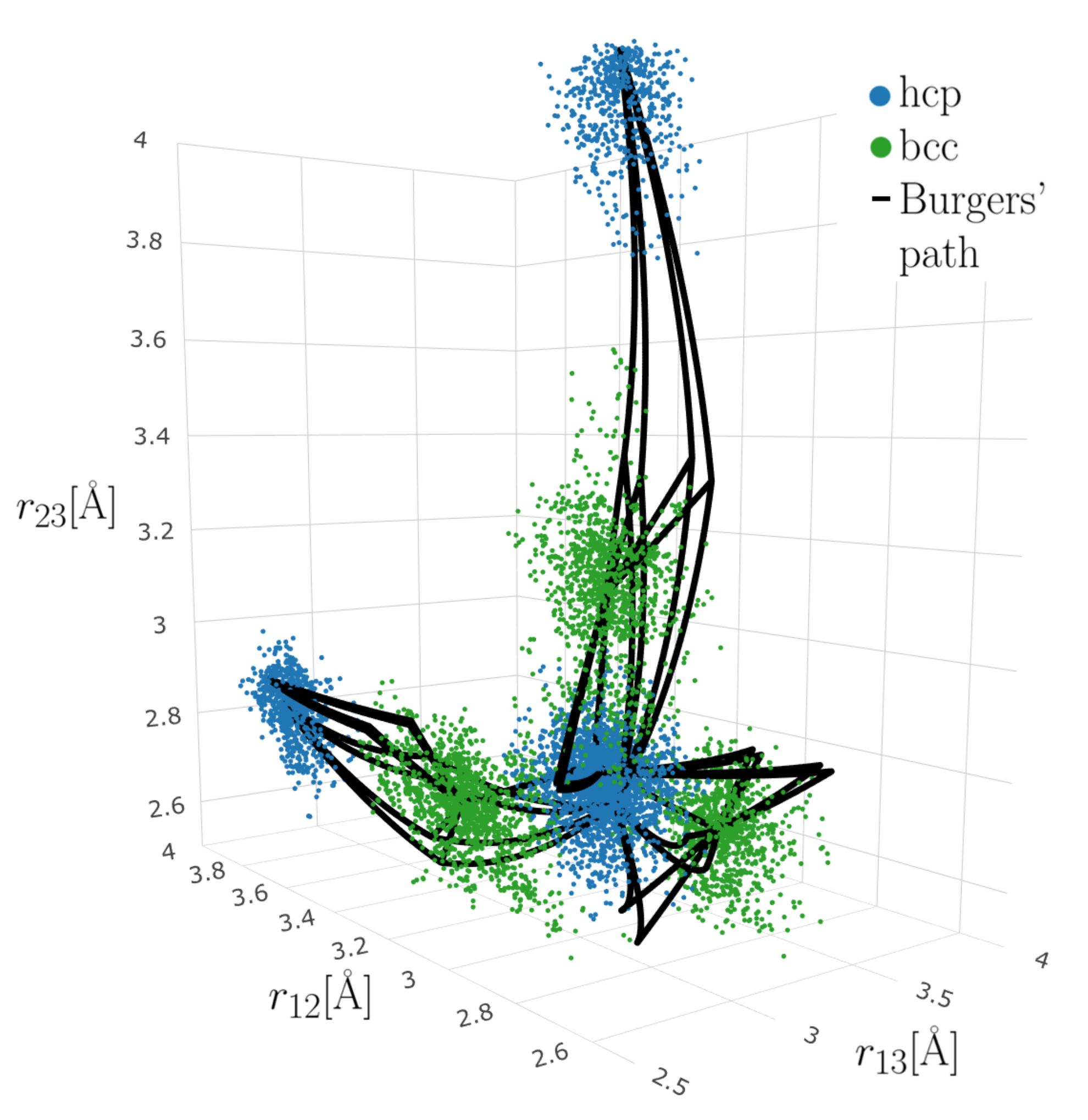}}
    \caption{Visualisation of the sparsity of the Ti training sets: each
    datapoint represents a 3-body cluster, described through interatomic
    distances $r_{12}, r_{13}, r_{23}$, contained in the primitive cell
    configurations of Set 1. The Burgers' path explores regions of 3-body space
    far from the regions with datapoints, which demonstrates that the Burgers'
    path test requires significant generalisation away from the training set.}
     \label{fig:3b_ti_burgers_path}
\end{figure}

To analyse this we use body order expansion to investigate the
Burgers' path test in a different way. By plotting the 3-body distances $r_{12}$, $r_{13}$, $r_{23}$ of the
primitive cell training database, Set 1,  we can visualise the clustering of
data in the $V_{3}$ space. Figure \ref{fig:3b_ti_burgers_path} shows the
clustering of data around the relaxed bcc/hcp configurations and shows the
Burgers' path connecting the two configurations as well. The training data was sampled
at a low temperature (T=100K) and we therefore see a limited
exploration of data away from the relaxed configurations. The Burgers' path
clearly explores configurations in this $V_{3}$ space where there is little to
no data present, showing that even in this low dimension this test can be
considered as a non-trivial generalisation.

We believe that the lack of accuracy along the Burgers' path of SOAP-GAP  model is due
to the lack of training data in that region. It is a manifestation of high
dimensional fits giving answers that are still regular, but have uncontrolled errors when extrapolating away from
the training database. The aPIP models perform
significantly better in this test because, as we propose, models fitted in low
dimension (e.g. the dimensionality of the body orders) will in general perform better in
generalisation tests compared to high dimensional fits such as SOAP-GAP  models.

It will also be interesting to systematically explore the relative benefits and possible unification of regularisation and active-learning style techniques that bring in new data in previously unexplored regions\cite{Nandi2019-bl, Podryabinkin2017-mo}.

\section{Conclusion}
The purpose of this paper was two-fold: Firstly, we developed {\em atomic PIPs}
(aPIPs), a generalisation of PIPs~\cite{Braams2009-wi}, interatomic potentials
constructed from permutation invariant polynomials for material systems by
applying the PIP construction to atomic body ordered expansions of the total
energy and endowing them with usual cutoff mechanisms. By fitting the polynomial coefficients to solid
training sets (rather than clusters in vacuum) we were able to obtain an
accuracy comparable with the SOAP-GAP models for tungsten~\cite{Szlachta2014-tb}
and silicon~\cite{Bartok:2018ih} (on a non-trivial subset of the full training
set) using low body-orders, four or five, which are still at least an order of
magnitude faster to evaluate than SOAP-GAP.

Secondly, we studied the generalisation (extrapolation) properties of the aPIPs.
We developed novel regularisation mechanisms that exploit the low-dimensional
structure of the body-ordered terms to ensure correct qualitative behaviour of
the potentials,  such as smoothness,  well away from the training set. We
showed that such a regularisation is crucial to achieve the
extrapolation properties of the Gaussian process based
SOAP-GAP~\cite{Szlachta2014-tb,Bartok:2018ih}. Indeed, our regularisation techniques
are amenable to fine-tuning which enabled us to significantly improve on the SOAP-GAP
model in several tests. Thus, we have established that our framework provides a novel
 ``tool kit'' for fitting interatomic potentials for materials with high accuracy and excellent
transferability, across a wide range of bonding chemistries.

The silicon, tungsten and titanium datasets are openly available at \url{http://www.libatoms.org/Home/DataRepository}, the aPIP framework can be found at \url{https://github.com/cortner/NBodyIPs.jl}.

\section{Acknowledgements}

GC and C. vd O. acknowledge the support of UKCP grant number EP/K014560/1. C. vd O. would like to acknowledge the support of EPSRC (Project Reference: 1971218) and Dassault Syst\`emes UK. CO and GD are supported by Leverhulme Research Project Grant RPG-2017-191.

\bibliography{biblio.bib}

\end{document}


\title{Supplementary Material: \\
Regularised Atomic Body-Ordered Permutation-Invariant Polynomials for the Construction of Interatomic Potentials
}

\author{Genevi\`eve Dusson$^*$\thanks{\texttt{g.dusson@warwick.ac.uk}}}
\affiliation{Mathematics Institute, University of Warwick, Coventry CV47AL, United Kingdom}

\author{Cas van der Oord$^*$\thanks{\texttt{casv2@eng.cam.ac.uk}}}
\affiliation{Engineering Laboratory,
University of Cambridge,
Trumpington Street,
Cambridge, CB2 1PZ,
United Kingdom}

\author{G\'abor Cs\'anyi\thanks{\texttt{gc121@cam.ac.uk}}}
\affiliation{Engineering Laboratory,
University of Cambridge,
Trumpington Street,
Cambridge, CB2 1PZ,
United Kingdom}

\author{Christoph Ortner\thanks{\texttt{c.ortner@warwick.ac.uk}}}
\affiliation{Mathematics Institute, University of Warwick, Coventry CV47AL, United Kingdom}

\date{\today}

\maketitle




\section{Systematic Convergence}
The potential and least squares parameters employed in Section~II.F of the main
text are displayed in Table~\ref{tbl:apip_params_convergence}.

\begin{table}
   {\bf RMSE Convergence Test Parameters} \\[1mm]
   {\it (a) Distance Transform and Cutoff} \\[1mm]
   \begin{tabular}{ccccc}
      \toprule[1pt]
                &    &     &       W             &   Si  \\
     invariants & body-order & $~~p~$ & $~r_{\rm cut}$ [\AA] & $~r_{\rm cut}$ [\AA] \\
     \midrule[0.5pt]
     D & 2 & 2 & 8.48 & 8.23 \\
     D & 3 & 3 & 7.12 & 7.52 \\
     D & 4 & 3 & 5.75 & 6.35 \\
     D & 5 & 4 & 5.20 & 5.17 \\
      \midrule[1pt]
     DA  & 2 & 2 & 8.48 & 8.23 \\
     DA  & 3 & 3 & 5.74 & 6.11 \\
     DA  & 4 & 3 & 5.20 & 4.94 \\
     DA  & 5 & 4 & 4.38 & 3.76 \\
     \bottomrule[1pt]
   \end{tabular}

   \medskip

   {\it (b) Least Squares Test Weights} \\[1mm]
   \begin{tabular}{ccc}
      \toprule[1pt]
       &            W             &   Si  \\
      data type  & ~weights & ~weights \\
      \midrule[1pt]
      $W_F$ & 1 & 1 \\
      $\#{\bf R} \times W_E$ & 100 & 300 \\
      $\#{\bf R} \times W_V$ & 1 & 1 \\
     \bottomrule[1pt]
   \end{tabular}
   \caption{(a) aPIP hyperparameters for the RMSE convergence tests described in
   \S~II.F of the main text. The parameter $p$ refers to the distance transform
   parameter.  \\
   (b) aPIP weight parameters arising in the least squares functional
   employed for the RMSE convergence tests described in
   \S~II.F of the main text. The weights $W_*$ may depend on ${\bf R}$, where
    $\#{\bf R} = M$ denotes the number of atoms in a configuration ${\bf R}$.}
   \label{tbl:apip_params_convergence}
\end{table}

\subsection{Extended Convergence tables: W}
We present in Tables~\ref{tab:W_conv_BA} and~\ref{tab:W_conv_BL} the systematic
convergence of the energy and forces together with the energy and forces
envaluation cost respectively with a distance-angle and a distance-based
potential.

\begin{table}[h!]
   \begin{tabular}{|l|l|l|l|l|l|l|l|l|} \hline
      \multicolumn{4}{|c|}{Degrees} &
                  \# basis
                    & \multicolumn{2}{|c|}{
                  \begin{tabular}{c} Timings \\ (ms/atom) \end{tabular}} & \multicolumn{2}{|c|}{RMSE} \\ \hhline{----~----}
                 2B & 3B & 4B & 5B &  functions  & energy & forces &
                 \begin{tabular}{c} energy \\ (meV) \end{tabular}
                  & \begin{tabular}{c} forces \\ (eV/A) \end{tabular}
                 \\ \Xhline{4\arrayrulewidth}8 & 0 & 0 & 0 & 8 & 0.0618 & 0.0704 & 31.7 & 0.445\\ \hline
      11 & 0 & 0 & 0 & 11 & 0.0703 & 0.0752 & 26.3 & 0.392\\ \hline
      14 & 0 & 0 & 0 & 14 & 0.065 & 0.0699 & 23.3 & 0.344\\ \hline
      17 & 0 & 0 & 0 & 17 & 0.0665 & 0.0713 & 20.7 & 0.333\\ \hline
      20 & 0 & 0 & 0 & 20 & 0.0663 & 0.0722 & 20.2 & 0.326\\ \hline
      20 & 5 & 0 & 0 & 42 & 0.184 & 0.257 & 13 & 0.256\\ \hline
      20 & 7 & 0 & 0 & 66 & 0.187 & 0.271 & 10.3 & 0.213\\ \hline
      20 & 9 & 0 & 0 & 103 & 0.204 & 0.307 & 8.62 & 0.176\\ \hline
      20 & 11 & 0 & 0 & 155 & 0.226 & 0.334 & 6.23 & 0.143\\ \hline
      20 & 13 & 0 & 0 & 225 & 0.265 & 0.383 & 5.26 & 0.129\\ \hline
      20 & 15 & 0 & 0 & 302 & 0.292 & 0.44 & 4.95 & 0.123\\ \hline
      20 & 17 & 0 & 0 & 349 & 0.31 & 0.512 & 4.83 & 0.121\\ \hline
      20 & 17 & 6 & 0 & 464 & 0.677 & 1.43 & 3.98 & 0.105\\ \hline
      20 & 17 & 8 & 0 & 696 & 0.851 & 1.72 & 3.28 & 0.0949\\ \hline
      20 & 17 & 10 & 0 & 1233 & 1.08 & 2.3 & 2.6 & 0.083\\ \hline
      20 & 17 & 12 & 0 & 2313 & 1.59 & 3.47 & 1.99 & 0.0703\\ \hline
      20 & 17 & 14 & 0 & 2842 & 1.9 & 4.41 & 1.85 & 0.0673\\ \hline
      20 & 17 & 14 & 6 & 2977 & 2.41 & 8.56 & 1.79 & 0.0657\\ \hline
      20 & 17 & 14 & 7 & 3113 & 2.46 & 8.56 & 1.74 & 0.0646\\ \hline
      20 & 17 & 14 & 8 & 3364 & 2.5 & 8.72 & 1.65 & 0.0626\\ \hline
      20 & 17 & 14 & 9 & 3815 & 2.64 & 9.04 & 1.55 & 0.0602\\ \hline
      20 & 17 & 14 & 10 & 4582 & 2.7 & 9.88 & 1.43 & 0.057\\ \hline
      20 & 17 & 14 & 11 & 5113 & 2.8 & 10.4 & 1.38 & 0.0555\\ \hline
      \end{tabular}
      \caption{Systematic convergence of the energy and forces for a tungsten database with distance-angle descriptors}
      \label{tab:W_conv_BA}
\end{table}

\begin{table}[h!]
   \begin{tabular}{|l|l|l|l|l|l|l|l|l|} \hline
      \multicolumn{4}{|c|}{Degrees} &
                  \# basis
                    & \multicolumn{2}{|c|}{
                  \begin{tabular}{c} Timings \\ (ms/atom) \end{tabular}} & \multicolumn{2}{|c|}{RMSE} \\ \hhline{----~----}
                 2B & 3B & 4B & 5B &  functions  & energy & forces &
                 \begin{tabular}{c} energy \\ (meV) \end{tabular}
                  & \begin{tabular}{c} forces \\ (eV/A) \end{tabular}
                 \\ \Xhline{4\arrayrulewidth}10 & 0 & 0 & 0 & 10 & 0.0631 & 0.0643 & 27.6 & 0.406\\ \hline
      12 & 0 & 0 & 0 & 12 & 0.0589 & 0.0616 & 25.7 & 0.389\\ \hline
      14 & 0 & 0 & 0 & 14 & 0.0624 & 0.0622 & 23.3 & 0.344\\ \hline
      16 & 0 & 0 & 0 & 16 & 0.0595 & 0.062 & 20.7 & 0.335\\ \hline
      18 & 0 & 0 & 0 & 18 & 0.0599 & 0.0641 & 20.5 & 0.329\\ \hline
      20 & 0 & 0 & 0 & 20 & 0.0632 & 0.0634 & 20.2 & 0.326\\ \hline
      20 & 5 & 0 & 0 & 35 & 0.218 & 0.262 & 13.6 & 0.26\\ \hline
      20 & 7 & 0 & 0 & 50 & 0.218 & 0.278 & 12.8 & 0.247\\ \hline
      20 & 9 & 0 & 0 & 72 & 0.221 & 0.275 & 10.9 & 0.223\\ \hline
      20 & 11 & 0 & 0 & 102 & 0.228 & 0.285 & 9.03 & 0.2\\ \hline
      20 & 13 & 0 & 0 & 142 & 0.239 & 0.3 & 8.42 & 0.188\\ \hline
      20 & 15 & 0 & 0 & 193 & 0.256 & 0.319 & 7.5 & 0.168\\ \hline
      20 & 17 & 0 & 0 & 256 & 0.27 & 0.347 & 6.43 & 0.146\\ \hline
      20 & 17 & 6 & 0 & 327 & 0.966 & 1.2 & 5.2 & 0.129\\ \hline
      20 & 17 & 8 & 0 & 450 & 0.993 & 1.27 & 4.41 & 0.117\\ \hline
      20 & 17 & 10 & 0 & 721 & 1.05 & 1.44 & 3.65 & 0.104\\ \hline
      20 & 17 & 12 & 0 & 1258 & 1.15 & 1.73 & 3.13 & 0.0954\\ \hline
      20 & 17 & 14 & 0 & 2240 & 1.37 & 2.14 & 2.61 & 0.0855\\ \hline
      20 & 17 & 14 & 6 & 2379 & 1.9 & 2.93 & 2.52 & 0.0839\\ \hline
      20 & 17 & 14 & 7 & 2528 & 1.85 & 2.91 & 2.47 & 0.0829\\ \hline
      20 & 17 & 14 & 8 & 2819 & 1.86 & 2.92 & 2.41 & 0.0817\\ \hline
      20 & 17 & 14 & 9 & 3358 & 1.88 & 2.92 & 2.32 & 0.0799\\ \hline
      20 & 17 & 14 & 10 & 4332 & 1.88 & 2.98 & 2.21 & 0.0777\\ \hline
      20 & 17 & 14 & 11 & 6023 & 1.92 & 2.98 & 2.05 & 0.0747\\ \hline
      \end{tabular}
       \caption{Systematic convergence of the energy and forces for a tungsten training set with distance-based descriptors}
      \label{tab:W_conv_BL}
\end{table}

\subsection{Extended Convergence tables: Si} 
We present in Tables~\ref{tab:Si_conv_BA} and~\ref{tab:Si_conv_BL} the systematic convergence of the energy and forces for a silicon training set with distance-angle and distance-based potentials. We also present the energy and forces envaluation costs for these potentials.

\begin{table}[h!]
   \begin{tabular}{|l|l|l|l|l|l|l|l|l|} \hline
      \multicolumn{4}{|c|}{Degrees} &
                  \# basis
                    & \multicolumn{2}{|c|}{
                  \begin{tabular}{c} Timings \\ (ms/atom) \end{tabular}} & \multicolumn{2}{|c|}{RMSE} \\ \hhline{----~----}
                 2B & 3B & 4B & 5B &  functions  & energy & forces &
                 \begin{tabular}{c} energy \\ (meV) \end{tabular}
                  & \begin{tabular}{c} forces \\ (eV/A) \end{tabular}
                 \\ \Xhline{4\arrayrulewidth}10 & 0 & 0 & 0 & 10 & 0.055 & 0.0589 & 350 & 1.1\\ \hline
      12 & 0 & 0 & 0 & 12 & 0.0554 & 0.0594 & 352 & 1.15\\ \hline
      14 & 0 & 0 & 0 & 14 & 0.0556 & 0.0606 & 321 & 1.2\\ \hline
      16 & 0 & 0 & 0 & 16 & 0.0563 & 0.0676 & 270 & 1.49\\ \hline
      18 & 0 & 0 & 0 & 18 & 0.0641 & 0.0658 & 239 & 1.27\\ \hline
      20 & 5 & 0 & 0 & 37 & 0.149 & 0.191 & 22.9 & 0.386\\ \hline
      20 & 7 & 0 & 0 & 57 & 0.141 & 0.194 & 14.3 & 0.314\\ \hline
      20 & 9 & 0 & 0 & 87 & 0.146 & 0.204 & 12.4 & 0.278\\ \hline
      20 & 11 & 0 & 0 & 129 & 0.16 & 0.216 & 9.93 & 0.268\\ \hline
      20 & 13 & 0 & 0 & 185 & 0.176 & 0.239 & 8.67 & 0.254\\ \hline
      20 & 15 & 0 & 0 & 257 & 0.195 & 0.263 & 7.92 & 0.258\\ \hline
      20 & 17 & 5 & 0 & 364 & 0.542 & 1.19 & 4.78 & 0.164\\ \hline
      20 & 17 & 7 & 0 & 456 & 0.591 & 1.29 & 3.9 & 0.148\\ \hline
      20 & 17 & 9 & 0 & 670 & 0.72 & 1.64 & 3.33 & 0.131\\ \hline
      20 & 17 & 11 & 0 & 1110 & 1 & 2.1 & 2.64 & 0.12\\ \hline
      20 & 17 & 13 & 0 & 1933 & 1.7 & 3.32 & 2.13 & 0.107\\ \hline
      20 & 17 & 14 & 6 & 2475 & 1.78 & 3.85 & 1.79 & 0.0965\\ \hline
      20 & 17 & 14 & 7 & 2545 & 1.78 & 3.84 & 1.73 & 0.0953\\ \hline
      20 & 17 & 14 & 8 & 2675 & 1.78 & 3.86 & 1.68 & 0.0935\\ \hline
      20 & 17 & 14 & 9 & 2880 & 1.79 & 3.84 & 1.64 & 0.0922\\ \hline
      20 & 17 & 14 & 10 & 3229 & 1.79 & 3.84 & 1.57 & 0.0907\\ \hline
      20 & 17 & 14 & 11 & 3759 & 1.78 & 3.91 & 1.47 & 0.0889\\ \hline
      \end{tabular}
   \caption{Systematic convergence of the energy and forces for a silicon database with distance-angle descriptors}
      \label{tab:Si_conv_BA}
\end{table}

\begin{table}
   \begin{tabular}{|l|l|l|l|l|l|l|l|l|} \hline
      \multicolumn{4}{|c|}{Degrees} &
                  \# basis
                    & \multicolumn{2}{|c|}{
                  \begin{tabular}{c} Timings \\ (ms/atom) \end{tabular}} & \multicolumn{2}{|c|}{RMSE} \\ \hhline{----~----}
                 2B & 3B & 4B & 5B &  functions  & energy & forces &
                 \begin{tabular}{c} energy \\ (meV) \end{tabular}
                  & \begin{tabular}{c} forces \\ (eV/A) \end{tabular}
                 \\ \Xhline{4\arrayrulewidth}10 & 0 & 0 & 0 & 10 & 0.054 & 0.0554 & 350 & 1.1\\ \hline
      12 & 0 & 0 & 0 & 12 & 0.0532 & 0.057 & 352 & 1.15\\ \hline
      14 & 0 & 0 & 0 & 14 & 0.0545 & 0.0563 & 321 & 1.2\\ \hline
      16 & 0 & 0 & 0 & 16 & 0.0547 & 0.0566 & 270 & 1.49\\ \hline
      18 & 0 & 0 & 0 & 18 & 0.0546 & 0.0574 & 239 & 1.27\\ \hline
      20 & 5 & 0 & 0 & 33 & 0.154 & 0.178 & 23.2 & 0.427\\ \hline
      20 & 7 & 0 & 0 & 48 & 0.157 & 0.189 & 19.2 & 0.325\\ \hline
      20 & 9 & 0 & 0 & 70 & 0.155 & 0.19 & 17 & 0.308\\ \hline
      20 & 11 & 0 & 0 & 100 & 0.163 & 0.196 & 13.9 & 0.308\\ \hline
      20 & 13 & 0 & 0 & 140 & 0.169 & 0.205 & 12.9 & 0.301\\ \hline
      20 & 15 & 0 & 0 & 191 & 0.179 & 0.212 & 11.3 & 0.295\\ \hline
      20 & 17 & 5 & 0 & 230 & 0.482 & 0.587 & 7.53 & 0.201\\ \hline
      20 & 17 & 7 & 0 & 310 & 0.489 & 0.602 & 6.14 & 0.184\\ \hline
      20 & 17 & 9 & 0 & 496 & 0.51 & 0.645 & 5.03 & 0.169\\ \hline
      20 & 17 & 11 & 0 & 880 & 0.545 & 0.756 & 4.13 & 0.154\\ \hline
      20 & 17 & 13 & 0 & 1613 & 0.609 & 0.88 & 3.15 & 0.141\\ \hline
      20 & 17 & 14 & 6 & 1752 & 1.28 & 2.24 & 3.1 & 0.14\\ \hline
      20 & 17 & 14 & 7 & 1901 & 1.19 & 2.25 & 3.05 & 0.139\\ \hline
      20 & 17 & 14 & 8 & 2192 & 1.2 & 2.26 & 2.98 & 0.137\\ \hline
      20 & 17 & 14 & 9 & 2731 & 1.23 & 2.3 & 2.85 & 0.134\\ \hline
      20 & 17 & 14 & 10 & 3705 & 1.27 & 2.37 & 2.69 & 0.13\\ \hline
      20 & 17 & 14 & 11 & 5396 & 1.31 & 2.51 & 2.49 & 0.126\\ \hline
      \end{tabular}
   \caption{Systematic convergence of the energy and forces for a silicon database with distance-based descriptors}
   \label{tab:Si_conv_BL}
\end{table}

\section{Regularised aPIP Fits}

\subsection{Weights}
The least squares weights in the regularised aPIP fits described in
\S~III.B of the main text are defined as follows: we scale energy and virial weights by
$(\#{\bf R})^{-1/2}$, which leads to
\begin{align}
   \notag
   W_{\rm E} &= \frac{w_{\rm E} \times w_{\rm cfg}}{\sqrt{\# \bf R}}, \\
   \label{eq:sqrt-weights}
   W_{\rm F} &= w_{\rm F} \times w_{\rm cfg}, \quad \text{and} \\
   \notag
   W_{\rm V} &= \frac{w_{\rm V} \times w_{\rm cfg}}{\sqrt{\# \bf R}},
\end{align}
where $w_{\rm cfg}$ is a configuration type dependent weight modifier which
are given in
Tables~\ref{tbl:W_reg} and~\ref{tbl:Si_reg}.

\subsection{Regularisation Parameters}
Tables~\ref{tbl:W_reg} and~\ref{tbl:Si_reg} show the detailed potential and
regularisation parameters for the unregularised and regularised aPIP fits
described in \S~III of the main text.

Each row of Tables~\ref{tbl:W_reg} and~\ref{tbl:Si_reg} describes a Laplace
regulariser in terms of the body-order $n$ to which it is applied, an interval
$r_{ij}  \in [r_0, r_1]$ over which the regulariser is applied, a regularisation
parameter $\gamma_n$, the number of Sobol
points used to construct the regulariser, and a distance transform for  the
distance variable $u_{ij} = u_{ij}(r_{ij})$. The Sobol points will then be
equidistributed in the $(u_{ij}, w_{ijk})$ coordinates. These may differ from
the transforms used to construct the potential itself in order to have maximal
flexibility where to concentrate the Sobol points. We also note that each
body-order $V_n$ may have more than one regulariser applied to it. In particular
we typically apply a second regulariser in the regime $r_{ij} < r_{\rm nn}$
where fewer training-points are available in order to better control
oscillations.

\begin{table}
   \begin{minipage}{\linewidth}
   {\bf W aPIP Parameters} \\[2mm]
   {\it (a) Distance Transform and Cutoff} \\[1mm]
   \begin{tabular}{cccc}
      \toprule[1pt]
      $~n~$ & $~p~$ & $~r_{\rm cut}~$ & degree \\
      \midrule[1pt]
      2 & 2 & 8.5 \AA & 17 \\
      3 & 3 & 6.3 \AA & 15 \\
      4 & 3 & 5.2 \AA & 13 \\
      \bottomrule[1pt]
   \end{tabular} \\[2mm]
   %
   %
   {\it (b) Least Squares Weights  \eqref{eq:sqrt-weights}} \\[1mm]
   \begin{tabular}{rcc}
      \toprule[1pt]
      configuration && $w_{\rm cfg}$ \\
      \midrule[1pt]
      unit cells && 5.0  \\
      bulk MD && 1.0     \\
      vacancy && 1.0     \\
      surface && 1.0     \\
      $\gamma$-surface && 0.2 \\
      $\gamma$-s. vacancy && 0.2 \\
      \bottomrule[1pt]
   \end{tabular}
   \quad
   \begin{tabular}{rcc}
      \toprule[1pt]
      weight && value\\
      \midrule[1pt]
      $w_E$ && 30.0  \\
      $w_F$ && 1.0 \\
      $w_V$ && 5.0 \\
      &&  \\
      &&  \\
      &&  \\
      \bottomrule[1pt]
   \end{tabular} \\[2mm]
   {\it (c) Regularisation Parameters}\\[1mm]
   \begin{tabular}{ccccc}
      \toprule[1pt]
     $n$ & $[r_0, r_1]/\text{\AA}$ & ~$\gamma_n$~ & npoints & transform \\
     \midrule[1pt]
     3  & [2.0, 6.3] & 0.3 & 3,000 & $(r_{\rm nn}/r)^{3}$ \\
     3  & [1.9, 2.33] & 0.3 & 500 & $r_{\rm nn}/r$ \\
     4  & [1.9, 5.2] & 0.3 & 10,000 & $r_{\rm nn}/r$ \\
     4  & [1.9, 2.2] & 0.3 & 10,000 & $r_{\rm nn}/r$ \\
     \bottomrule[1pt]
   \end{tabular}
   \end{minipage}
   \caption{Parameters for W aPIP for regularisation and extrapolation tests.}
   \label{tbl:W_reg}
\end{table}

\begin{table}
   \begin{minipage}{\linewidth}
   {\bf Si aPIP Parameters} \\[2mm]
   {\it (a) Distance Transform and Cutoff} \\[1mm]
   \begin{tabular}{cccc}
      \toprule[1pt]
      $~n~$ & $~p~$ & $~r_{\rm cut}~$ & degree \\
      \midrule[1pt]
      2 & 2 & 8.7 \AA & 17 \\
      3 & 3 & 6.2 \AA & 15 \\
      4 & 3 & 5.4 \AA & 13 \\
      5 & 3 & 4.2 \AA & 11 \\
      \bottomrule[1pt]
   \end{tabular} \\[2mm]
   {\it (b) Least Squares Weights \eqref{eq:sqrt-weights}} \\[1mm]
   \begin{tabular}{rcc}
      \toprule[1pt]
      configuration && $w_{\rm cfg}$ \\
      \midrule[1pt]
      dia         && 1.0 \\
      amorph      && 1.0 \\
      bt          && 1.0 \\
      vacancy     && 1.0 \\
      sp2         && 1.0 \\
      surface110  && 1.0 \\
      surface111  && 1.0 \\
      surface001  && 1.0 \\
      \bottomrule[1pt]
   \end{tabular}  \quad
   \begin{tabular}{rcc}
      \toprule[1pt]
      observation && weight\\
      \midrule[1pt]
      $w_E$ && 100.0 \\
      $w_F$ && 1.0 \\
      $w_V$ && 2.0 \\
      && \\
      && \\
      && \\
      && \\
      && \\
      \bottomrule[1pt]
   \end{tabular} \\[2mm]
   {\it (c) Regularisation Parameters}\\[1mm]
   \begin{tabular}{ccccc}
      \toprule[1pt]
     $n$ & $[r_0, r_1]/\text{\AA}$ & $\gamma_n$ & npoints & transform \\
     \midrule[1pt]
     2  & [1.8, 8.7] & 0.025 & 300 & $r_{\rm nn}/r$ \\
     3  & [1.8, 2.2] & 0.01 & 1,000 & $r$ \\
     3  & [1.8, 6.3] & 0.03 & 1,000 & $(r_{\rm nn}/r)^{2}$ \\
     4  & [1.8, 2.2] & 0.03 & 10,000 & $r$ \\
     4  & [1.8, 5.4] & 0.03 & 10,000 & $(r_{\rm nn}/r)^{2}$ \\
     5  & [1.8, 2.2] & 0.2 & 30,000 & $r$ \\
     5  & [1.8, 4.3] & 0.2 & 10,000 & $(r_{\rm nn}/r)^{2}$ \\
     \bottomrule[1pt]
   \end{tabular}
   \end{minipage}
   \caption{Parameters for Si aPIP for regularisation and extrapolation tests.}
   \label{tbl:Si_reg}
\end{table}

\subsection{Sequential Fitting: W}
The procedure to produce the regularised aPIP fit to the W training set is as follows:
\begin{enumerate}
   \item The 1-body term is given by $V_1(R) = E_0$, where $E_0 = -9.194835$ eV.
   \item We fit $V_2$ only to the unit cells, bulk MD, vacancy and dislocation training subset, and replace the virial weight 3.0 with 70.0. We add a laplace regulariser on the interval $[1.37\textrm{ \AA}, 8.5\textrm{ \AA}]$ with 1000 regularisation points. To carry out the fit we use the rrQR algorithm with truncation parameter $10^{-9}$.
   \item The repulsive core potential is chosen with spline point $r_{\rm S} = 2.192$ and $e_\infty = 0.0$.
   \item We then fit the $V_3$ and $V_4$ terms jointly using the full training set with parameters as described in Table~\ref{tbl:W_reg}, using the standard QR algorithm.
\end{enumerate}

\subsection{Sequential Fitting: Si}
The procedure to produce the regularised aPIP fit to the Si training set is as follows:
\begin{enumerate}
   \item The 1-body term is given by $V_1(R) = E_0$, where $E_0 = -158.545$ eV.
   \item We fit $V_2, V_3, V_4$ jointly with parameters specified by Table~\ref{tbl:Si_reg}, observing that the Si training set already ensures
    a smooth dimer-shape for the $V_2$ term. There was no regularisation setting that produced a dimer shape for $V_2$ when fitted separately.
   \item The repulsive core potential is chosen with spline point $r_{\rm S} = 2.02 {\rm \AA}$ and $e_\infty = -2.0$.
   \item The modification of Step 2 affects the fit accuracy on the {\rm bt} training subset. We therefore refit $V_3, V_4$ with $V_1, V_2$ as a reference potential and
         parameters specified in Table~\ref{tbl:Si_reg}.
   \item Finally we fit $V_5$ with $V_1, V_2, V_3, V_4$ as a reference potential
      and parameters specified by Table~\ref{tbl:Si_reg}.
\end{enumerate}

\subsection{Details of Ti Fits}
In this section we describe the fits to the titanium training sets described in
\S~IV.C in the main document. For Set 1 we fit both unregularised and
regularised aPIP potentials (aPIP$_1$ (unreg) and aPIP$_1$ (reg)) as well as a SOAP-GAP potential.
For Set 2 we only fit an unregularised aPIP potential (aPIP$_2$ (unreg)).

The least square weights and model parameters for the aPIP fits are described in
detail in Table~\ref{tbl:apip_ti_params}.
The parameters for the SOAP-GAP fit are given in Table~\ref{tbl:gap_ti_params}

\begin{table}
   \begin{minipage}{\linewidth}
   {\bf Ti aPIP Parameters} \\[2mm]
   {\it (a) Distance Transform and Cutoff} \\[1mm]
   \begin{tabular}{cccc}
      \toprule[1pt]
      $~n~$ & $~p~$ & $~r_{\rm cut}~$ & degree \\
      \midrule[1pt]
      2 & 2 & 8.5 \AA & 25 \\
      3 & 3 & 7.53 \AA & 20 \\
      4 & 2 & 6.08 \AA & 10 \\
      \bottomrule[1pt]
   \end{tabular} \\[2mm]
   {\it (b) Least Squares Weights  \eqref{eq:sqrt-weights}} \\[1mm]
   \begin{tabular}{rcc}
      \toprule[1pt]
      configuration && $w_{\rm cfg}$ \\
      \midrule[1pt]
      bcc (acc) && 1.0  \\
      bcc (rej) && 1.0     \\
      hcp (acc) && 1.0     \\
      hcp (rej) && 1.0     \\
      bcc FDC* && 125.0 \\
      hcp FDC* && 50.0 \\
      \bottomrule[1pt]
   \end{tabular}
   \quad
   \begin{tabular}{rcc}
      \toprule[1pt]
      weight && value\\
      \midrule[1pt]
      $w_E$ && 100.0  \\
      $w_F$ && 1.0 \\
      &&  \\
      &&  \\
      &&  \\
      &&  \\
      \bottomrule[1pt]
   \end{tabular} \\[2mm]
   {\it (c) Regularisation Parameters}\\[1mm]
   \begin{tabular}{ccccc}
      \toprule[1pt]
     $n$ & $[r_0, r_1]/\text{\AA}$ & $c_{\rm reg}$ & npoints & transform \\
     \midrule[1pt]
     2  & [4.0, 6.0] & 0.005 & 100 & $(r_{\rm nn}/r)^{3}$ \\
     3  & [2.0, 4.0] & 0.5 & 1000 & $r_{\rm nn}/r$ \\
     4  & [2.0, 6.0] & 0.05 & 10,000 & $r_{\rm nn}/r$ \\
     \bottomrule[1pt]
   \end{tabular}
   \end{minipage}
   \caption{Parameters for Ti aPIP for regularisation and extrapolation tests. (*) FDC is Finite Displacement Configuration, referring to a 3x3x3 supercell with a single atom displaced which was added to predict the phonon spectrum accurately. }
   \label{tbl:apip_ti_params}
\end{table}

\begin{table}
        {\bf Ti GAP Parameters} \\[2mm]
        \begin{tabular}{rcc}
      \toprule[1pt]
      Parameter && Value \\
      \midrule[1pt]
      $\sigma_{atom}$         && 0.5 \AA \\
      $\sigma^{energy}_{\nu}$         && 0.001 eV \\
      $\sigma^{force}_{\nu}$         && 1$\%$ of $f_{\nu}$  \\
      $n_{max}$ && 10 \\
      $l_{max}$ && 10 \\
      $\zeta$ && 4 \\
      $n_{\rm sparse}$ && 900 \\
      \bottomrule[1pt]
   \end{tabular}
   \caption{Parameters for Ti GAP model. $\sigma^{force}_{\nu}$ was chosen to be a small
   percentage of the force per atom $f_{\nu}$ in order to accurately fit the phonon spectra  }
   \label{tbl:gap_ti_params}
\end{table}

\clearpage